\documentclass[review, 1p]{elsarticle}

\usepackage{lineno,hyperref}
\modulolinenumbers[5]

\journal{Journal of \LaTeX\ Templates}

\usepackage{url}
\usepackage[utf8]{inputenc}
\usepackage{tabularx}

\usepackage{orcidlink}
\usepackage{graphicx}
\usepackage{dcolumn}
\usepackage{bm}
\usepackage[utf8]{inputenc}
\usepackage[T1]{fontenc}
\usepackage{mathptmx}
\usepackage{etoolbox}
\usepackage{xcolor}
\usepackage{graphicx}
\usepackage{lineno,hyperref}
\usepackage{blindtext}
\usepackage{epsfig}
\usepackage{latexsym}
\usepackage{graphics}
\usepackage{epsfig}
\usepackage{amsmath,amssymb,amsfonts,theorem,color,bm}
\usepackage{multirow, multicol, boldline}
\usepackage{stfloats} 
\usepackage{float}
\usepackage{subfig}
\usepackage{soul}
\usepackage{caption}
\hypersetup{colorlinks,allcolors=black}
\usepackage[normalem]{ulem}

\newcommand\etal{{\it et al. }}
\newcommand\ie{{\it i.e. }}

\newcommand{\lineExpVz}   {\protect\raisebox{0.5pt}{\textcolor{black}{--{\tiny$\blacktriangle$}--{\tiny$\blacktriangle$}--}}}  
\newcommand{\lineExpVx}   {\protect\raisebox{0.5pt}{\textcolor{black}{--$\bullet$--$\bullet$--}}}               

\newcommand{\lineSimVz}   {\protect\raisebox{0.5pt}{\textcolor{blue}{------}}}                     
\newcommand{\lineSimVx}   {\protect\raisebox{0.5pt}{\textcolor{orange}{------}}}                             

\newcommand{\lineOurVz}   {\protect\raisebox{0.5pt}{\textcolor{red}{------}}}                      
\newcommand{\lineOurVx}   {\protect\raisebox{0.5pt}{\textcolor{green}{------}}}  

\begin{document}

\begin{frontmatter}

\title{Characterizing Intraventricular Flow Patterns via Modal Decomposition Techniques in Idealized Left Ventricle Models.}

\newcommand{\orcidEL}{0000-0002-4514-6471}
\newcommand{\orcidJGM}{0000-0002-7422-5320}
\newcommand{\orcidSLCM}{0000-0003-3605-7351}
\newcommand{\orcidMN}{0000-0002-5582-6905}

\author[addressUPM]{E. Lazpita \orcidlink{\orcidEL}}
\cortext[mycorrespondingauthor]{Corresponding author}
\ead{e.lazpita@upm.es}

\author[addressRWTH]{M. Neidlin \orcidlink{\orcidMN}}
\author[addressUPM,addressCSC]{J. Garicano-Mena \orcidlink{\orcidJGM}}
\author[addressUPM,addressCSC]{S. Le Clainche \orcidlink{\orcidSLCM}}

\address[addressUPM]{ETSI Aeron\'autica y del Espacio - Universidad Polit\'ecnica de Madrid, 28040 Madrid, Spain}
\address[addressCSC]{Center for Computational Simulation (CCS), 28660 Boadilla del Monte, Spain}
\address[addressRWTH]{Department of Cardiovascular Engineering, Institute of Applied Medical Engineering, Medical Faculty, RWTH Aachen University, 52074 Aachen, Germany}

\begin{abstract}

Understanding the formation, propagation, and breakdown of the main vortex ring (VR) is essential for characterizing left ventricular (LV) hemodynamics, as its dynamics have been linked to the onset and progression of cardiovascular diseases. In this study, two idealized LV geometries, a semi-ellipsoidal chamber and a more rounded configuration, are analyzed using computational fluid dynamics (CFD) simulations under physiological conditions, with the aim of investigating the fluid mechanisms that govern VR evolution during diastole. Modal decomposition techniques, specifically proper orthogonal decomposition (POD) and higher order dynamic mode decomposition (HODMD), are employed to identify dominant coherent structures and track their temporal behavior.

To the authors’ knowledge, this is the first time such an analysis is conducted with the explicit goal of unraveling the physics of vortex ring dynamics in idealized ventricular chambers. The comparative approach reveals that geometric morphology plays a central role in modulating the flow: in one case, early interaction between the VR and the ventricular wall, driven by the chamber’s shape, triggers strong nonlinear interactions and a more intricate dynamic evolution. In the other, the vortex ring propagates more freely toward the apex before dissipating, resulting in a more organized flow pattern and simpler spectral content.

These findings advance the understanding of flow-based indicators relevant to early diagnosis and treatment planning in cardiovascular disease. Moreover, they illustrate how the choice of ventricular geometry can influence not only the simulated hemodynamics, but also the effectiveness of data-driven analysis tools, depending on the clinical context under study.

\end{abstract}


\end{frontmatter}

\section{\label{sec:Introduction} Introduction}

Cardiovascular health is of paramount importance, as heart-related diseases remain among the leading causes of morbidity and mortality worldwide. In 2019 alone, an estimated 17.9 million people died from cardiovascular diseases (CVDs), accounting for 32\% of all global deaths. Of these, 85\% were due to heart attacks and strokes \cite{WHO2024}. Gaining insight into the internal flow dynamics of the heart is crucial for the diagnosis, monitoring, and treatment of such conditions.

In particular, the analysis of blood flow within the left ventricle (LV) is essential for assessing cardiac pumping efficiency and identifying potential functional abnormalities. During early diastole, the inflow through the mitral valve forms a vortex ring (VR) that travels toward the apex. This VR not only facilitates efficient ventricular filling but also aids systolic ejection by conserving angular momentum. Its formation, evolution, and stability are governed by anatomical and physiological conditions, and its characterization is considered a potential indicator of cardiac health \cite{Pedrizzetti2014Vortex}.

\begin{figure}[h]
    \centering
    \includegraphics[width=\textwidth]{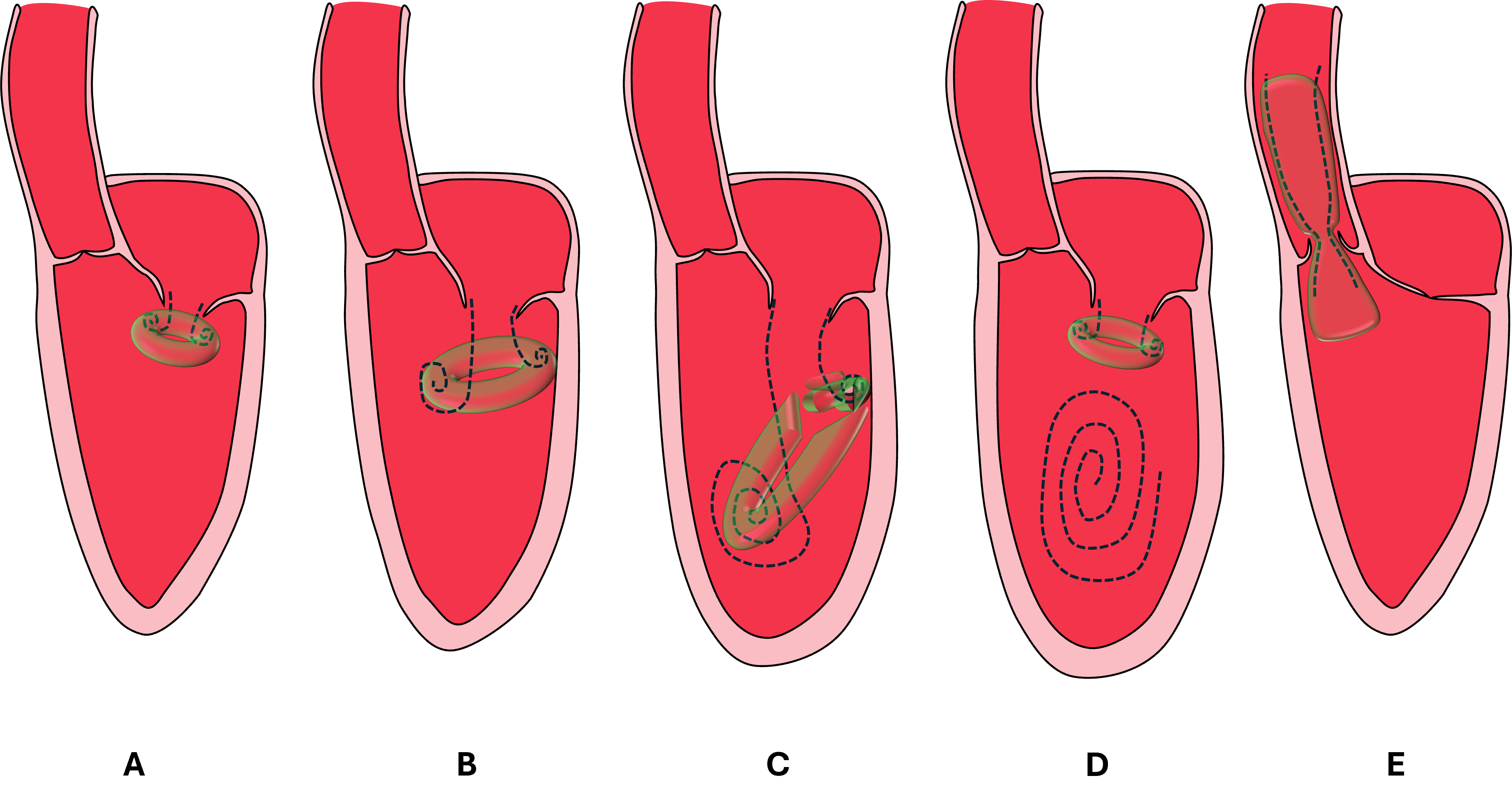}
    \caption{Qualitative schematic of VR dynamics during the cardiac cycle: (A) early diastole, (B) peak E-wave velocity, (C) velocity deceleration, (D) A-wave inflow, and (E) systolic ejection.}
    \label{fig:flowfeatures}
\end{figure}

The dynamics of this intraventricular vortex ring have been widely studied via experimental and numerical approaches \cite{Toger2012Vortex, Le2012Vortex}. The presence of the VR is key for efficient diastolic filling, minimizing energy dissipation, and promoting effective intraventricular transport. These dynamics are sketched in Fig. \ref{fig:flowfeatures}.

At the onset of diastole (A), the mitral valve opens, allowing blood to enter the LV. This peak in velocity inflow is denoted as E-wave (B) and generates a vortex ring at the mitral annulus, initially circular in shape, which propagates toward the apex. As the ring travels, interaction with the surrounding ventricular walls induces asymmetries that deform it into an elliptical configuration.

Subsequently, the VR interacts with nearby walls (C), generating secondary vortex tubes that wrap around the primary structure. These interactions promote the growth of instabilities within the core, which, by the end of diastole, result in the breakdown of the main vortex into smaller-scale structures.

A secondary, weaker vortex ring appears during the A-wave in late diastole (D), formed by a second inflow through the mitral valve. However, due to the lower velocity and volume of this inflow, the secondary ring is less energetic and has a limited propagation range.

During systole (E), the ventricle contracts, increasing internal pressure and opening the aortic valve, driving blood into the systemic circulation. This final phase generates a strong outflow, completing the cardiac cycle. 

These dynamic mechanisms are further analyzed and illustrated in this study for two distinct LV geometries. Understanding the main vortex ring, its formation, evolution, and dissipation, is vital for characterizing LV hemodynamics. Due to its close link to ventricular efficiency, deviations from the typical vortex behavior may signal impaired cardiac function.

Intraventricular blood flow is inherently unsteady and complex, and its numerical simulation via computational fluid dynamics (CFD) has become increasingly prevalent. CFD allows for the investigation of pulsatile flow, vortex dynamics, shear stresses, and pressure gradients throughout the cardiac cycle.

In recent years, the application of CFD to cardiac modeling has experienced a significant increase. For instance, Grunwald \etal \cite{Grunwald2022Intraventricular} introduced a methodology for simulating patient-specific LV geometries. Korte \etal \cite{Korte2023Hemodynamic} adopted a similar approach to study mitral insufficiency under exercise conditions. Other studies have pushed these methods further, such as Ref. \cite{Fedele2023Comprehensive}, which presents a comprehensive electromechanical model of the human heart including accurate geometry, myocardial fiber orientation, active force generation models, and  coupled with a zero-dimensional closed-loop lumped parameters circulatory system model. An even more advanced multiphysics and multiscale geometric framework was developed in Ref. \cite{Zingaro2024Electromechanics}, enabling detailed modeling of full-heart electromechanics and hemodynamics.

The CFD data generated from intraventricular simulations is typically high-dimensional and time-resolved, posing challenges for interpretation and analysis. Traditional flow visualization techniques, such as streamlines, velocity contour plots, or the \textit{Q-criterion}, are commonly used to identify coherent structures and qualitatively analyze flow behavior during the cardiac cycle \cite{Schenkel2009MRI, Nguyen2015Patient, Colorado2022Patient}. These methods are often complemented with experimental validation techniques such as particle image velocimetry (PIV) \cite{Fortini2013Three, DiLabbio2022Braids}.

However, these qualitative tools often fall short of revealing the intrinsic mechanisms driving flow evolution, frequencies, and flow instabilitites associated to the vortex breakdown. To address this, modal decomposition techniques such as proper orthogonal decomposition (POD) \cite{Berkooz1993Proper} and dynamic mode decomposition (DMD) \cite{Schmid2010Dynamic} have been increasingly applied to cardiovascular flows, particularly in arteries and veins, to extract coherent structures and dominant flow features \cite{Kazemi2022Reduced, Habibi2020Data}. Nevertheless, their application within the left ventricle remains limited. In a recent study, Wu \etal\ \cite{Wu2023Flow} demonstrated that the combination of Shake-the-Box and POD analysis offers a promising framework for investigating the complex, three-dimensional, and time-resolved nature of left ventricular flows. Similarly, Borja \etal\ \cite{Borja2024Deriving} applied POD to patient-specific left ventricular simulations, revealing characteristic modal patterns and quantifying the kinetic energy associated with each mode.

Moreover, the use of these techniques has grown significantly in recent years for the analysis of medical data. In the work by Groun \etal\ \cite{Groun2022Higher}, higher-order dynamic mode decomposition (HODMD) \cite{LeClainche2017Higher} was employed to analyze medical imaging and identify flow patterns associated with cardiovascular diseases. This information was subsequently integrated with deep learning techniques to develop an automatic classification system, which achieved substantially higher accuracy than models relying solely on image-based features \cite{Bell2025Automatic}. This shows the effectiveness to identify flow patterns in cardiac flows.

HODMD has previously been employed to identify instabilities and flow patterns in a variety of complex problems, including turbulent flows \cite{Corrochano2020Flow, Lazpita2022Generation, Munoz2022Topology}, as well as to develop reduced-order models (ROMs) \cite{LeClainche2018Reduced, Beltran2022Adaptive}. In this study, we present the results obtained using POD and HODMD, two widely used complementary techniques for extracting coherent structures from unsteady flow fields. The aim is to provide a detailed analysis of the formation and evolution of the main VR, and to identify the characteristic frequencies associated with its generation and the underlying instabilities.

While POD has been previously applied to the analysis of left ventricular flow in both experimental and computational studies~\cite{Borja2024Deriving, Wu2023Flow}, these applications have mainly focused on identifying energetic flow structures without explicitly linking them to the underlying temporal dynamics.
In contrast, the present study, to the best of the authors’ knowledge, is the first to employ HODMD to analyze the formation and destabilization mechanisms of the LV vortex ring of intraventricular flows. HODMD offers a key advantage over POD in that each dynamic mode is associated with a single frequency, enabling a finer decomposition of the flow’s temporal dynamics. In contrast, POD provides spatially orthogonal modes whose temporal evolution may contain contributions from multiple frequencies, which can complicate physical interpretation in unsteady or turbulent flows. However, in simpler laminar regimes, individual POD modes often align closely with dominant frequencies, and their structure may closely resemble that of DMD modes. The strength of HODMD lies in its ability to isolate temporally orthogonal modes with well-defined frequencies, making it particularly suitable for identifying flow instabilities and transitions such as vortex breakdown. This frequency-based decomposition also enables the design of targeted control strategies aimed at modifying specific modes to suppress or enhance certain flow behaviors. By comparing two idealized ventricular geometries, this study highlights how geometric variations influence vortex dynamics, offering valuable insight for the selection of modeling strategies in future studies, particularly when investigating pathological conditions.

This paper is organized as follows: Section~\ref{sec:Database} presents the two idealized left ventricle models, offers a description of the CFD simulation process and comments on the validation. Section~\ref{sec:Methods} introduces the modal decomposition techniques applied in the analysis, POD and HODMD, that are in turn used in Section~\ref{sec:Results} to gain insight on the flow dynamics on both configurations. Finally, Section~\ref{sec:Conclusions} summarizes the key conclusions of this work.

\section{\label{sec:Database} Databases and Validation}
The idealized models presented in this study serve as a reference framework for investigating intraventricular hemodynamics. The main objective is to analyze key flow characteristics, with particular emphasis on the formation and evolution of the VR. These simplified geometries reduce computational complexity while providing greater control over the simulations \cite{Tagliabue2017Complex}. 

The first model, to which we shall refer to as \texttt{Ideal 1}, is a semi-ellipsoidal approximation of the left ventricle, derived from Ref. \cite{Zheng2012Computational}. It has been instrumental in establishing our CFD setup, as extensively detailed in Ref. \cite{Lazpita2024Modeling}. The second model, \texttt{Ideal 2}, is based on a left ventricle geometry obtained from the numerical and experimental study described in Vedula \etal \cite{Vedula2014Computational}. This model provides an additional validation source, offering a clearer simulation environment due to its lower ejection fraction and a larger ventricular chamber, which in turn enhances the visibility of flow structures during diastole \cite{Kjeldsberg2023Verified}.

In the validation and description of the two simulation databases, we employ two key flow descriptors: the total kinetic energy (TKE) and the Q-criterion.

The TKE quantifies the global dynamic content of the velocity field at a given instant and is defined at each non-dimensional time instant \( t^* \) as:
\begin{equation}
    \mathrm{TKE}(t^*) = \int_{V(t^*)} \frac{1}{2}\rho \mathbf{v}^2\, dV,
\end{equation}
where \( V(t^*) \) is the time-dependent ventricular volume, \( \rho \) is the fluid density, and \( \mathbf{v} \) is the velocity vector. This quantity is commonly used to assess the energy content and flow intensity within the left ventricle throughout the cardiac cycle.

The Q-criterion \cite{Hunt1988Turbulent} is a scalar quantity that enables the identification of vortical structures by comparing the relative magnitude of rotation to strain in the velocity gradient tensor. It is defined as:

\begin{equation}
    Q = \frac{1}{2} \left( \|\bm{ \Omega }\|^2 - \|\mathbf{S}\|^2 \right),
\end{equation}
where $\mathbf{S} = \frac{1}{2} \left( \nabla \mathbf{u} + (\nabla \mathbf{u})^\top \right)$ is the symmetric part of the velocity gradient tensor (\ie strain rate tensor), and $\bm{\Omega} = \frac{1}{2} \left( \nabla \mathbf{u} - (\nabla \mathbf{u})^\top \right)$ is the antisymmetric part (vorticity tensor). Positive values of $Q$ indicate regions where the rotational effects dominate over the strain, making it a valuable tool to visualize coherent vortex structures, such as the vortex ring.

Throughout the analysis, we employ the dimensionless time variable $t^* = t/T$, where $T$ denotes the period of the cardiac cycle ($t^* \in [0, 1]$). In this study, a heart rate of 60 beats per minute (bpm) is adopted, yielding a period of \( T = 1\,\text{s} \). For the \texttt{Ideal 1} model, the cardiac cycle is divided into diastole, occurring in the interval $t^* \in [0, 0.67]$, and systole in $t^* \in [0.67, 1]$. In contrast, the \texttt{Ideal 2} model features a longer diastolic phase, with $t^* \in [0, 0.8]$, followed by a shorter systolic phase $t^* \in [0.8, 1]$.The timing of the transition from diastole to systole occurs when the maximum volume is achieved and is extracted from the literature, specifically from Zheng \etal~\cite{Zheng2012Computational} and Vedula \etal~\cite{Vedula2014Computational}.

\subsection{\label{subsec:ideal1-validation} Ideal geometry based on a semi-ellipsoid: \texttt{Ideal 1}}

\begin{figure}[ht]
    \centering
    \includegraphics[trim=0 0 0 0, clip, width=0.9\linewidth]{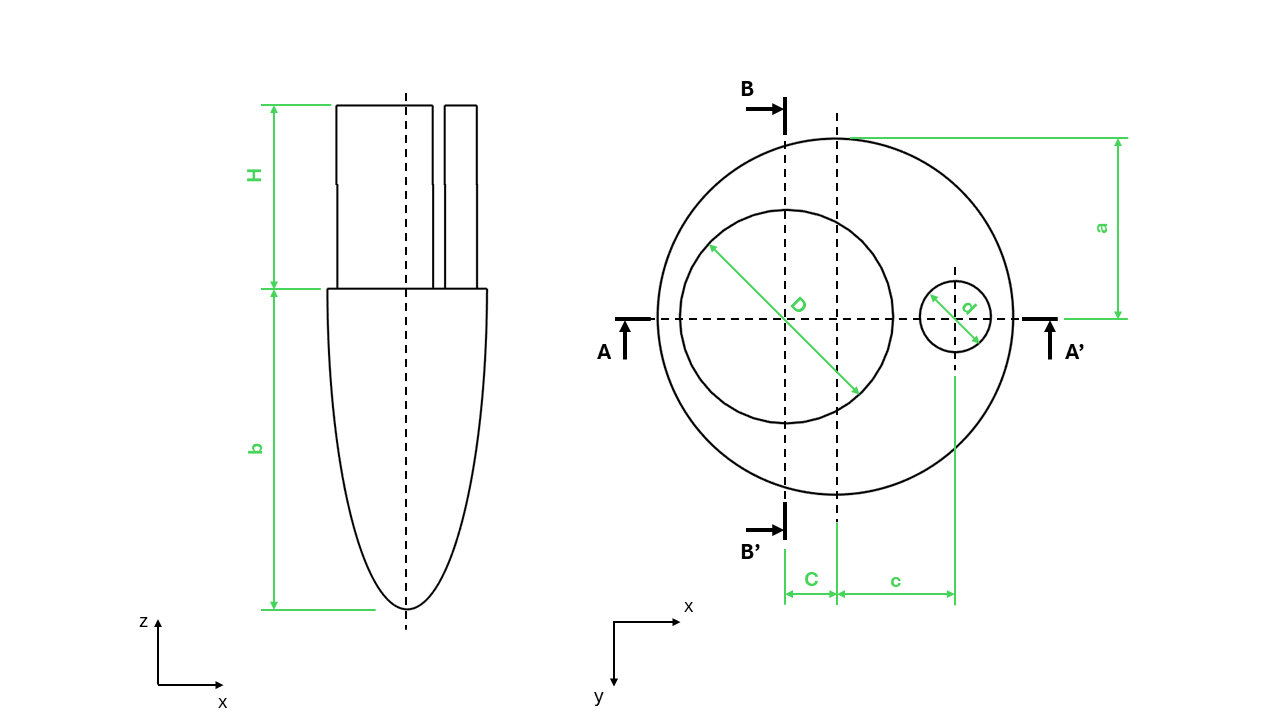}
    \caption{Geometrical scheme of the model \texttt{Ideal 1}, where the A-A' and B-B' planes are represented.}
    \label{fig:ideal1-geom}
\end{figure}

The \texttt{Ideal 1} model is represented as a semi-ellipsoid with two cylindrical tubes attached to its upper plane. The selection of geometrical parameters is crucial to ensure that the model accurately reflects physiological conditions, particularly the end-systolic volume (ESV), which corresponds to the minimum volume of the LV at the end of systole. A schematic representation of these parameters and their spatial locations is provided in Fig.~\ref{fig:ideal1-geom}. To characterize the model, we define the following non-dimensional geometric parameters, using the short semiaxis \( a \) of the ellipsoid as the reference length:
\begin{itemize}
    \item Semiaxis ratio: \( b/a = 4 \).
    \item Tube diameters: \( D/a = 1.2 \) (inlet), \( d/a = 0.4 \) (outlet).
    \item Tube heights: \( H/a = 2.4 \) for both inlet and outlet.
    \item Tube center offsets from the ellipsoid axis: \( C/a = 0.275 \) (inlet), \( c/a = 0.675 \) (outlet).
\end{itemize}
To achieve physiological values for the ventricular volume, the short semiaxis \( a \) should be set to 2 cm. The figure also illustrates the definition of two representative planes used throughout our analysis: the x-z symmetry plane at $y = 0$ (denoted as plane A-A') and the y-z plane intersecting the inlet axis along the $y$ direction (denoted as plane B-B'). Additionally, the sketch includes the definitions of the x, y, and z directions, which serve as reference axes for interpreting velocity and vorticity components (e.g., $v_x$ refers to the velocity component in the X direction, while $\omega_y$ denotes the vorticity component in the Y direction).

Accurately defining the fluid properties and governing parameters is crucial for obtaining meaningful simulation results in hemodynamic studies. While blood exhibits shear-thinning properties at lower shear rates, within the left ventricle (LV) and under physiological conditions, treating it as Newtonian is a reasonable simplification. Based on previous works \cite{Korte2023Hemodynamic}, we adopt a constant density of \(\rho = 1060\) $kg/m^3$ and a dynamic viscosity of \(\mu = 0.004\) $Pa \cdot s$.

To characterize the flow regime, we compute the Reynolds number (\(\mathrm{Re}\)), which represents the ratio of inertial to viscous forces. Using the inlet tube diameter, \(D\), as the characteristic length and the peak inlet velocity (\(V_{\text{max}} = 0.86\)) of the cardiac cycle, the Reynolds number is estimated as:  
\begin{equation}
    \mathrm{Re} = \frac{\rho V_{\text{max}} D}{\mu} \approx 5500 \, .
\end{equation}

As shown in Ref. \cite{He2022Numerical, Tagliabue2017Complex}, it is possible to accurately and efficiently capture the formation and evolution of the vortex ring by assuming that the flow behaves as laminar and not considering turbulence in the simulations. Therefore, as our objective is to capture the VR, in our study the blood flow is modelled as laminar.

In previous studies \cite{Lazpita2024ECCOMAS, Lazpita2024Modeling}, we validated our simulation framework by comparing two independent CFD solvers: Ansys Fluent \cite{Fluent} and Star-CCM+ \cite{Starccm}. The aim was to evaluate the influence of solver-specific numerical schemes and meshing strategies on the fidelity and reproducibility of the results. Simulations were conducted under identical boundary conditions and prescribed wall motion, based on a synthetic ventricular deformation model implemented in both solvers. Each solver underwent an independent mesh convergence study to ensure numerical accuracy. Following this analysis, \textit{Ansys Fluent} was selected as the reference solver for subsequent work, so that both ventricular models share a consistent computational methodology throughout the database generation.

\begin{figure}[h]
    \centering
    \includegraphics[width=0.75\textwidth]{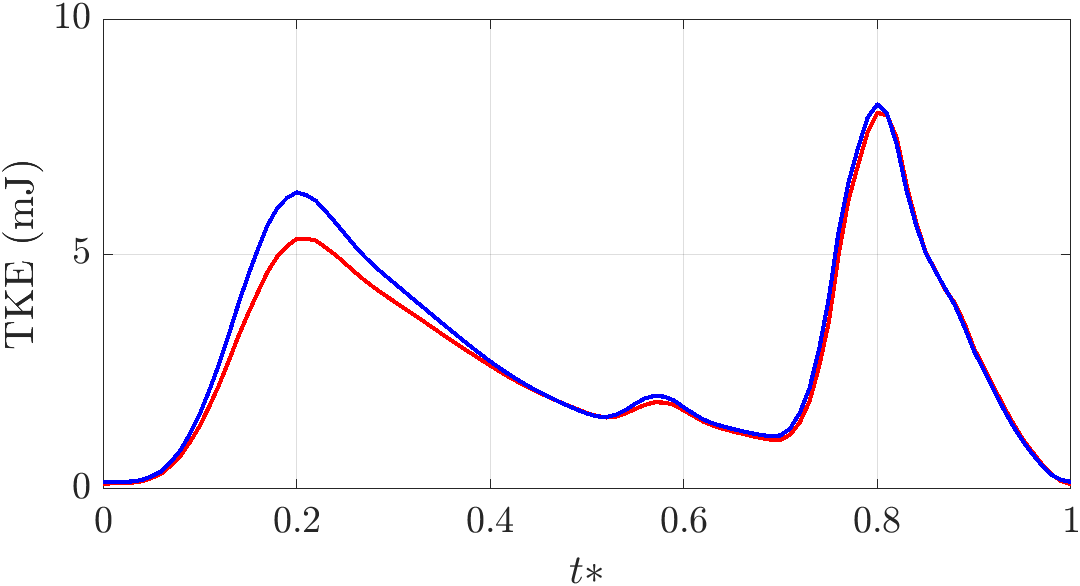}
    \caption{Comparison of TKE evolution over one cycle between results obtained from each boundary condition selection: velocity inlet/mass flow rate outlet~(\lineOurVz) \cite{Lazpita2024Modeling}, and pressure inlet/pressure outlet~(\lineSimVz).}
    \label{fig:ideal1-validation-mesh}
\end{figure}

In this work, we further assess the impact of different inlet/outlet boundary condition (BC) configurations. Specifically, we compare two alternatives: one prescribing a velocity profile at the inlet and a mass flow rate at the outlet (extracted from Ref.~\cite{Lazpita2024ECCOMAS,Lazpita2024Modeling}), and another imposing pressure BCs at both boundaries, closer to the setup used for the second model presented later (extracted from Ref.~\cite{Vedula2014Computational, Grunwald2022Intraventricular}). Figure~\ref{fig:ideal1-validation-mesh} shows the evolution of TKE over the cardiac cycle for both cases, using converged mesh resolutions. The TKE trends are overall consistent, with only minor deviations observed near the early diastolic peak. The RRMSE between the two configurations, computed according to Eq.~\ref{eq:RRMSE}, remains below 3\% throughout most of the cycle. A larger discrepancy appears during the E-wave of diastole, but it remains below 10\% at all times. While the solutions are not identical, as expected due to the distinct boundary condition formulations, they exhibit a satisfactory level of agreement. In particular, key flow features such as vortex ring formation, remain consistent across both cases. This supports the validity of either boundary condition configuration for capturing the main intraventricular dynamics analyzed in this study. Based on these results, we retain the pressure-based BC configuration for the remainder of this work.

\begin{figure}[h]
    \centering
    \includegraphics[width=\textwidth]{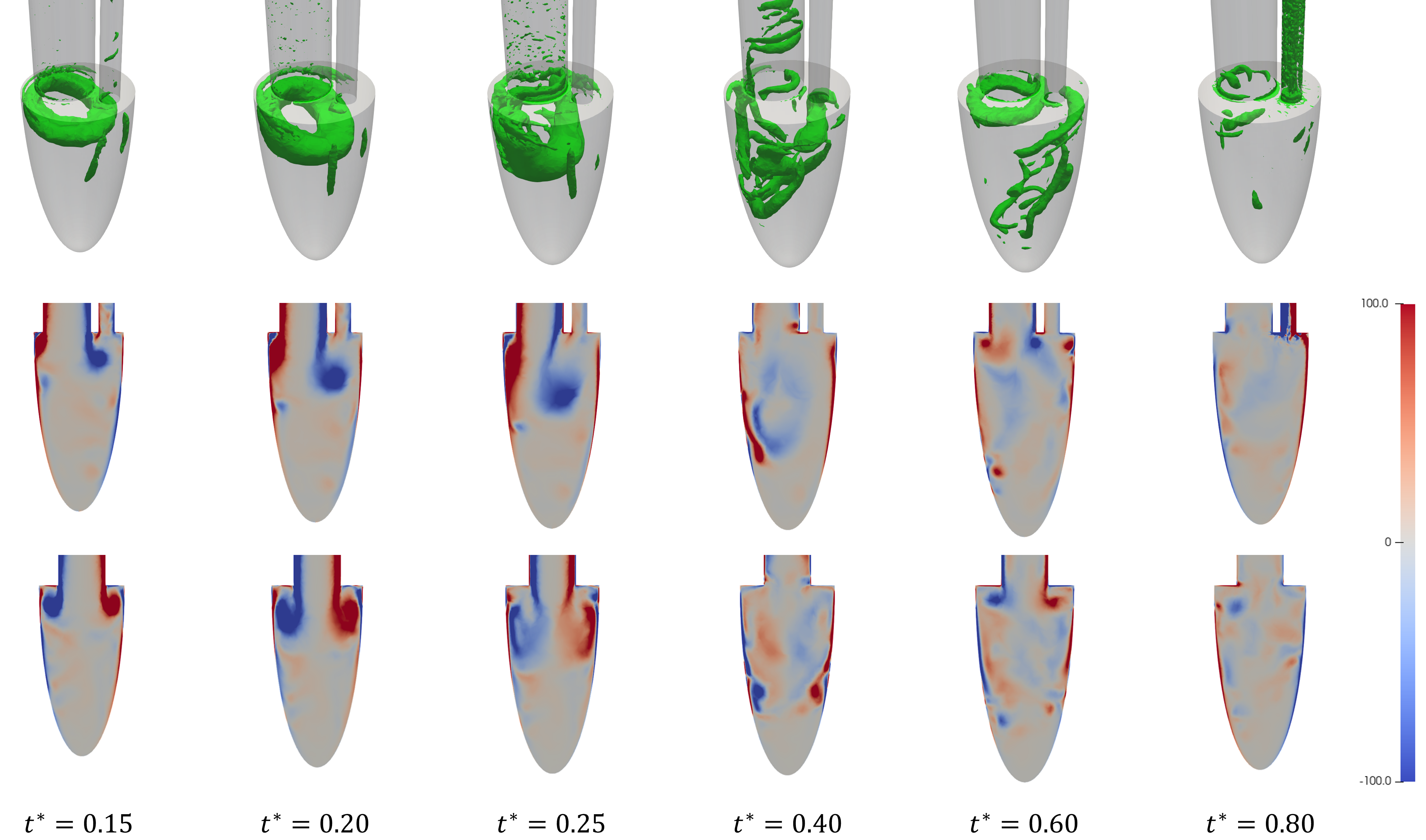}
    \caption{Temporal evolution of the VR in the \texttt{Ideal 1} database using: (top) Q-Criterion isocontours, (middle) $\omega_y$ vorticity contours on plane A-A' (Fig. \ref{fig:ideal1-geom}), and (bottom) $\omega_x$ vorticity contours on plane B-B' (Fig. \ref{fig:ideal1-geom}).}
    \label{fig:ideal1-flowfeatures}
\end{figure}

Figure \ref{fig:ideal1-flowfeatures} illustrates the evolution of the Q-Criterion isosurfaces traced at level 1000 and plane-normal vorticity during a heart cycle to characterize intraventricular flow structures. At the early diastolic phase ($t^*=0.15$), the vortex ring emerges at the exit of the inlet tube, corresponding to the mitral valve opening during the E-wave. Initially, the VR maintains a circular configuration. As it progresses into the ventricle, at $t^*=0.20$, the ring tilts due to proximity to the ventricular wall and the subsequent reduction in inflow velocity. This interaction leads to the ring adopting an elliptical shape upon impingement with the wall. By $t^*=0.25$, secondary vortex tubes develop due to the interaction between the vortex ring and the ventricular walls. These trailing vortex structures wrap around the primary VR by $t^*=0.40$ that contribute to its gradual dissipation. At $t^*=0.60$, the vortex ring is compleltely dissipated; however, immediately after a secondary, smaller vortex emerges at the mitral valve due to the A-wave. Since the main vortex ring rapidly dissipates, this secondary ring does not interact with it and develops for a longer time. Finally, after the mitral valve closes, ventricular contraction drives blood towards the aorta, creating a strong outflow during the late systolic phase, as observed at $t^*=0.80$. The results obtained exhibit strong agreement between each solver and align closely with the findings in Ref. \cite{Zheng2012Computational} (see Fig. 3).

As observed in Fig.~\ref{fig:ideal1-flowfeatures}, the vortex ring breaks well before reaching the apex. The early vortex breakup phenomenon is promoted by the relatively narrow chamber geometry: the strong geometrical constrains on the vortex evolution triggers the presence of secondary flow structures that interact with the main vortex and are connected to the non-linearities of the flow dynamics, leading to increased flow complexity. More details are presented in Section~\ref{sec:Results}.

To further analyze the vortex dynamics within the left ventricle, we examined two-dimensional contours of plane-normal vorticity. The second row presents these contours in the symmetry plane (Fig. \ref{fig:ideal1-geom}, plane A-A') of the idealized geometry. The asymmetry of the swirling motion in the vortex ring is evident during its formation and subsequent tilting at the early time points. The latter time points illustrate the recirculating flow following vortex dissipation and the emergence of the secondary ring at $t^*=0.60$. The final instant highlights the strong outflow during systole. Furthermore, bottom row shows vorticity contours in a plane that intersects the inlet tube along its axis (Fig. \ref{fig:ideal1-geom}, plane B-B'). This visualization emphasizes the initial vortex ring symmetry in the $y$ direction during formation and evolution until dissipation. The evolution of the VR observed in this study is consistent with the findings reported in Ref.~\cite{Zheng2012Computational}, where a similar dynamic behavior was described.

\subsection{\label{subsec:ideal2-validation} Ideal geometry based on experimental setup: \texttt{Ideal 2}}

Similar to \texttt{Ideal 1}, this model incorporates inlet and outlet tubes to represent the mitral and aortic valves. However, unlike the previous simplified geometry, the ventricular chamber in this case is not a straightforward semi-ellipsoid. Instead, the myocardial wall is derived from an experimental setup developed in Vedula \etal \cite{Vedula2014Computational}. The computational LV model was reconstructed from biplanar imaging data, where ventricular wall was captured from two perpendicular perspectives of the experimental setup and then processed to generate the three-dimensional geometry.

\begin{figure}[ht]
    \centering
    \includegraphics[trim=0 0 0 0, clip, width=0.9\linewidth]{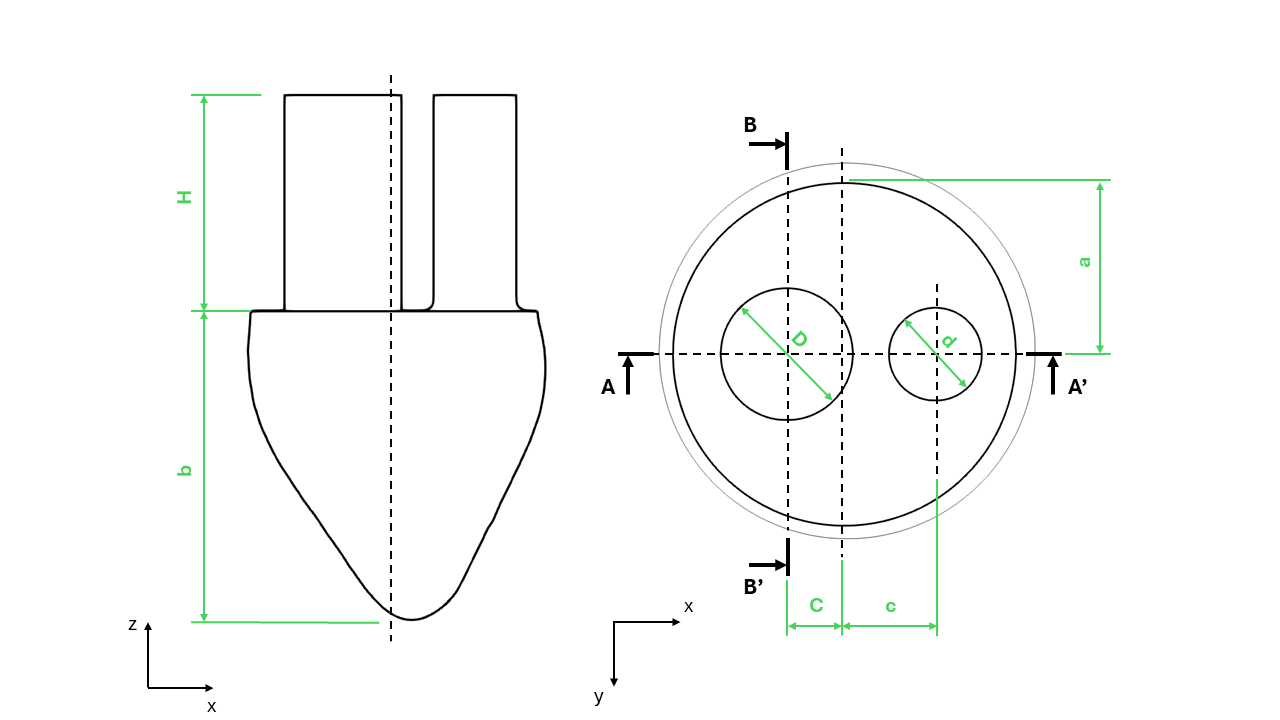}
    \caption{Same as Fig.~\ref{fig:ideal1-geom} for the \texttt{Ideal 2} model.}
    \label{fig:ideal2-geom}
\end{figure}

Although this model is based on experimental data, some geometrical features have been extracted from the idealized LV model to facilitate interpretation and comparison. Figure~\ref{fig:ideal2-geom} presents a schematic representation of the model, including the key non-dimensional geometric parameters, all scaled by the radius \( a \) of the base plane:
\begin{itemize}
    \item Chamber length ratio: \( b/a = 2.2 \).
    \item Tube diameters: \( D/a = 0.85 \) (inlet), \( d/a = 0.6 \) (outlet).
    \item Tube heights: \( H/a = 2.7 \) for both inlet and outlet.
    \item Tube center offsets from the main axis: \( C/a = 0.35 \) (inlet), \( c/a = 0.6 \) (outlet).
\end{itemize}
To replicate physiological ventricular volumes, the base radius \( a \) is set to 2.9 cm.
This figure again shows the two representative planes used throughout our analysis for this case: the x-z symmetry plane at $y = 0$ (denoted as plane A-A') and the y-z plane intersecting the inlet axis along the $y$ direction (denoted as plane B-B'). The definitions of the x, y, and z directions follows the same convention to the previous model.

The fluid configuration mirrors that of the \texttt{Ideal 1} model is employed, simplifying blood again as an incompressible Newtonian fluid with a constant density of $\rho = 1060$~kg/m$^3$ and a dynamic viscosity of $\mu = 0.004$~Pa$\cdot$s. For this LV model, the Reynolds number ($\mathrm{Re}$) is computed using the inlet tube diameter, $D$, as the characteristic length scale and the peak inlet velocity of the cardiac cycle, $V_{\text{max}} = 0.77$~m/s. This yields an estimated Reynolds number of $\mathrm{Re} \approx 4900$.

In order to replicate as closely as possible the reference study by Vedula \etal~\cite{Vedula2014Computational}, we impose pressure inlet and outlet boundary conditions and model valve dynamics by switching these boundaries to stationary walls at appropriate time intervals during the cardiac cycle. Since this model is introduced for the first time in this manuscript, we include the necessary mesh independence and validation results to ensure the reliability of the simulations.

\begin{figure}[h]
    \centering
    \includegraphics[width=0.45\textwidth]{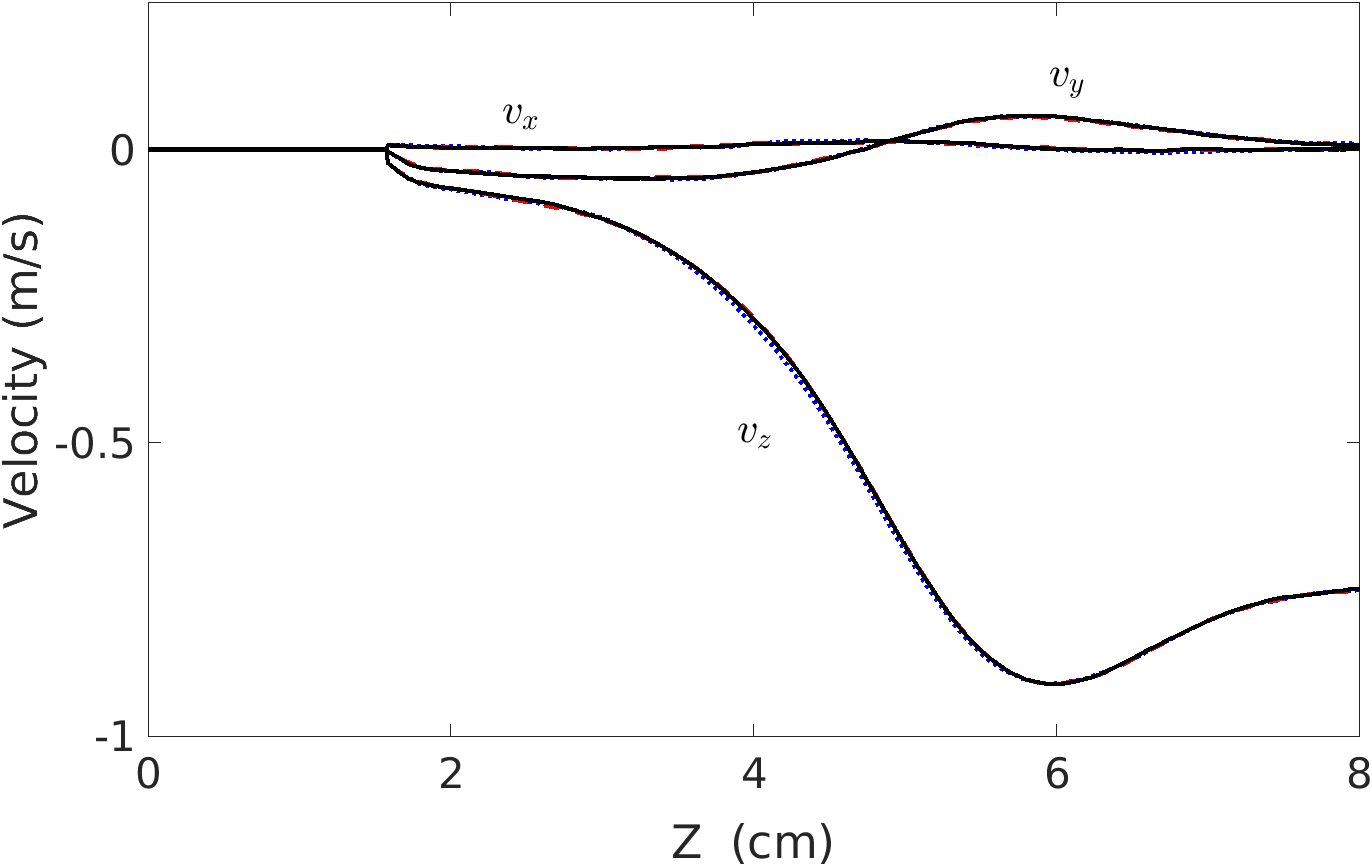}
    \includegraphics[width=0.45\textwidth]{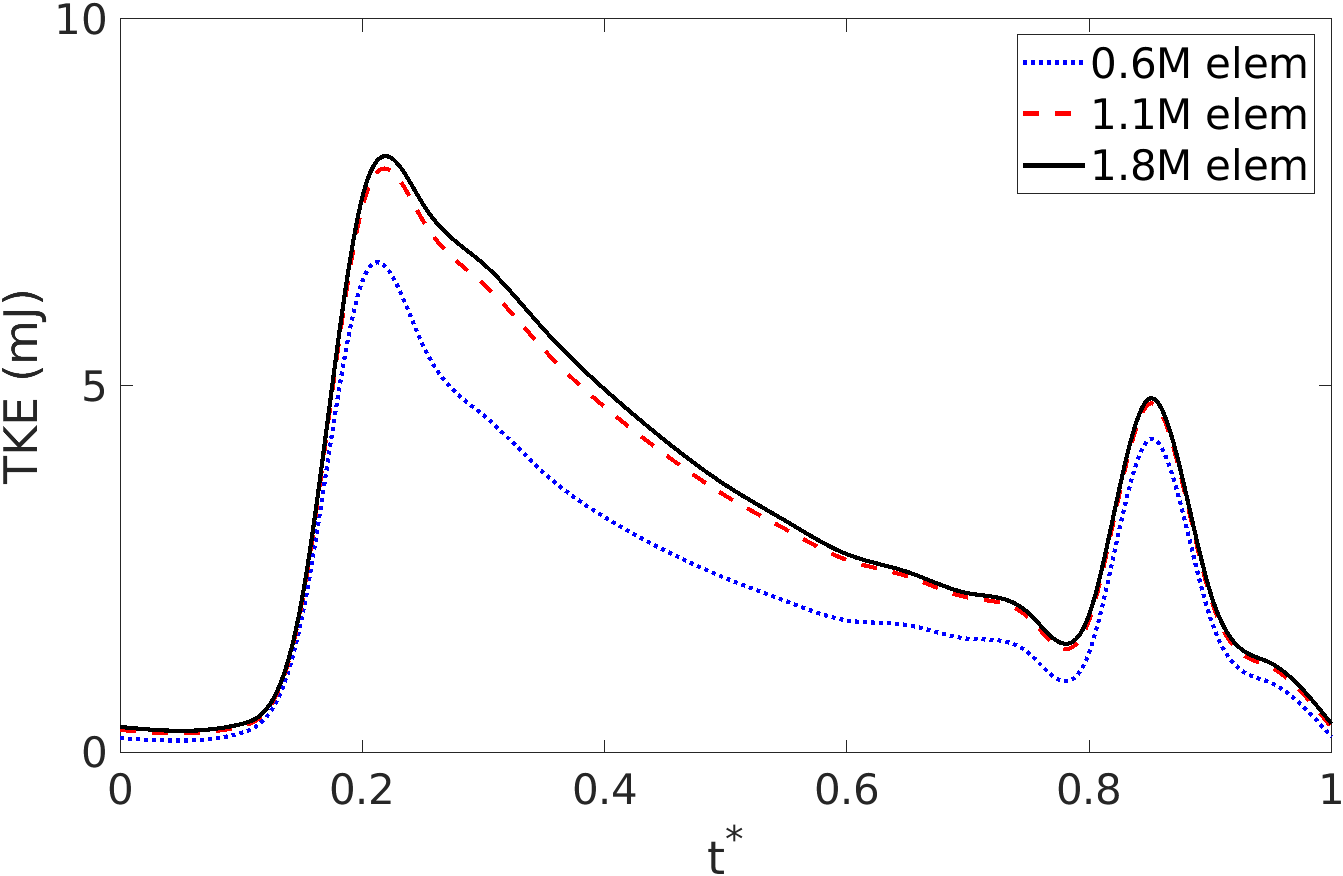}
    \caption{Comparison of mesh configurations: (left) velocity components along the mitral orifice centerline at $t^* = 0.20$, (right) TKE evolution over one cycle.}
    \label{fig:ideal2-validation-mesh}
\end{figure}

We perform a convergence study using three mesh sizes: coarse (0.6 M elements), medium (1.1 M elements) and fine (1.8 M elements), where "M" represents million of units.  Figure \ref{fig:ideal2-validation-mesh} presents the evolution of velocity components along the centerline of the mitral orifice at $t^* = 0.20$ and the TKE within the left ventricle over one cardiac cycle for each mesh configuration. The results show strong agreement between the medium and fine meshes, while the coarse mesh lacks sufficient resolution to capture key flow features. Notably, the TKE plot shows the closest alignment in diastole, with minor discrepancies during the rapid ventricular contraction in systole. Table~\ref{tab:ideal2-validation} reports the relative root mean square error (RRMSE) of each mesh with respect to the finest resolution, defined as:
\begin{equation}
    RRMSE = \frac{||L - L_{\text{true}}||}{||L_{\text{true}}||} \cdot 100 \, ,
    \label{eq:RRMSE}
\end{equation}
where \( L_{\text{true}} \) denotes the reference data from the finest mesh, and \( L \) represents the solution obtained from the coarse or medium mesh. The results demonstrate that the medium mesh maintains errors below 5\% across the most relevant variables. Thus, this configuration is chosen as the optimal balance between accuracy and computational cost.

\begin{table}[h]
    \centering
    \begin{tabular}{| l | c | c | c | c | c |}
        \hline
        \textbf{Mesh}  & \textbf{Elements} & $RRMSE(v_z)$ & $RRMSE(TKE)$  \\
        \hline
        Coarse     & 0.6 M  & 0.85\%  & 27.65\% \\
        Medium     & 1.1 M  & 0.25\%  &  3.50\% \\
        Fine       & 1.8 M  & -     & -     \\
        \hline
    \end{tabular}
    \caption{Mesh characteristics and relative error of velocity component \(v_z\) and TKE profiles for the \texttt{Ideal 2} model.}
    \label{tab:ideal2-validation}
\end{table}

Comparing our mesh configurations with those in Ref.~\cite{Vedula2014Computational} is not straightforward, as their methodology embeds the ventricle within a cartesian mesh, whereas in our case, the ventricle itself defines the control volume. To facilitate a fair comparison, we define the estimated average mesh density \( \psi \) as the ratio between the chamber volume (in mL) and the number of mesh elements. 

While the control volume in the reference study remains constant throughout the cardiac cycle, our simulations involve a deformable mesh, leading to time-varying volumes. Therefore, we report mesh densities at ESV and also at end diastolic volume (EDV). Table~\ref{tab:ideal2-mesh} presents the corresponding values for each mesh resolution.

As a general trend, the mesh density at ESV is more comparable to that reported in the reference study, Ref.~\cite{Vedula2014Computational}. At EDV, however, the density decreases due to the larger chamber size, particularly for the fine mesh configuration. This discrepancy becomes more pronounced when comparing to the fine mesh in Ref.~\cite{Vedula2014Computational}, which contains approximately 33 million elements, resulting in a mesh density of \( \psi = 0.20 \times 10^{-4} \text{ mL/elem} \). In contrast, our fine mesh yields densities of \( 0.80 \times 10^{-4} \) at ESV and \( 1.10 \times 10^{-4} \) at EDV.

Nevertheless, the velocity and TKE profiles shown in the previous figures indicate that numerical convergence is achieved with our current mesh sizes, suggesting that further mesh refinement offers marginal gains in accuracy. Thus, we conclude that our mesh densities are comparable and sufficient for capturing the main flow features in line with the reference study.

\begin{table}[h]
    \centering
    \begin{tabular}{| l | c | c | c | c | c |}
        \hline
        \multirow{2}{*}{Mesh size}  & \multicolumn{3}{c |}{Estimated Average Mesh Density [$\times 10^{-4}$ mL/elem]} \\
        \cline{2-4}
                & Vedula \etal \cite{Vedula2014Computational} & Current Study (ESV) & Current Study (EDV) \\
        \hline
        Coarse  & 3.05         & 2.65          & 3.65          \\
        Medium  & 1.30         & 1.35          & 1.90          \\
        Fine    & 0.20         & 0.80          & 1.10          \\
        \hline
    \end{tabular}
    \caption{Comparison of estimated average mesh density between our studies and Ref. \cite{Vedula2014Computational} for the \texttt{Ideal 2} model.}
    \label{tab:ideal2-mesh}
\end{table}

\begin{figure}[h]
    \centering
    \includegraphics[trim=0 0 250 0, clip, width=0.8\textwidth]{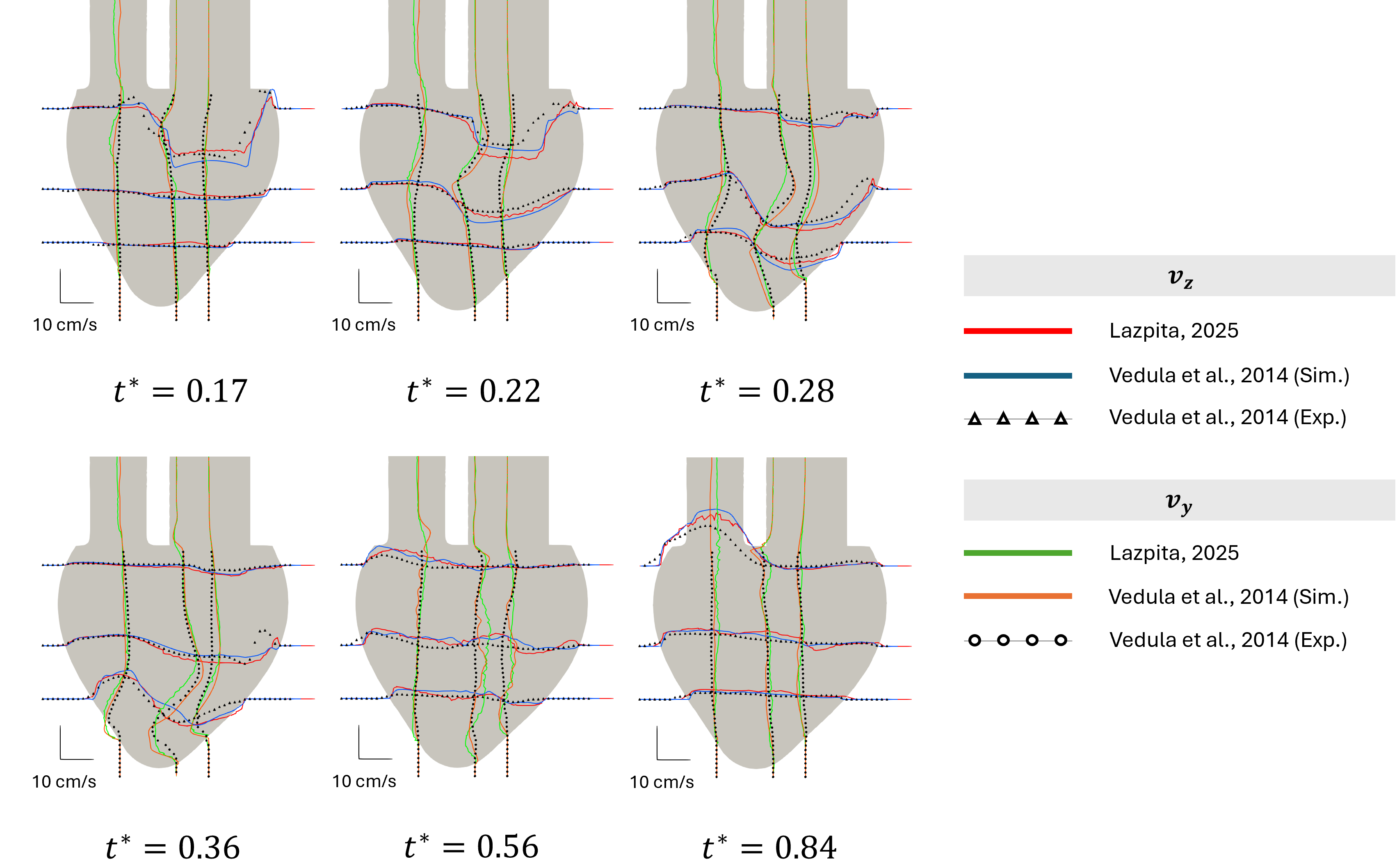}
    \caption{Velocity profiles along six lines in the A--A' plane at different time instants, comparing current simulations with data from \cite{Vedula2014Computational} - Fig. 6. 
    Reference experiments: \(v_z\)~(\lineExpVz) and \(v_x\)~(\lineExpVx); 
    Reference simulations: \(v_z\)~(\lineSimVz) and \(v_x\)~(\lineSimVx); 
    Current study: \(v_z\)~(\lineOurVz) and \(v_x\)~(\lineOurVx).}
    \label{fig:ideal2-validation-vel}
\end{figure}

We further compare our simulation results with both computational and experimental data (see Fig.~6 from Ref.~\cite{Vedula2014Computational}). Figure~\ref{fig:ideal2-validation-vel} presents velocity profiles along six lines in the symmetry plane: three horizontal lines for the \(v_z\) component and three vertical lines for the \(v_y\) component, evaluated at six representative time instants throughout the cardiac cycle (e.g., vortex ring formation, tilting, outflow). Velocity magnitudes can be quantitatively assessed by comparison with the reference lines plotted at \(\pm10\) cm/s. At early times, a strong inflow dominates the dynamics, resulting in velocity fields primarily aligned with the vertical direction. As the vortex ring develops and begins to recirculate, the flow acquires a noticeable component along the horizontal direction, particularly evident at \( t^* = 0.22 \), as the jet penetrates deeper into the cavity. This redirection of momentum continues during the filling phase, with the flow gradually turning toward the outlet axis. In the final stages of the cycle, the ventricular contraction drives a strong upward flow, realigning the velocity field with the vertical direction. Overall, our simulations show good agreement with both numerical and experimental references, supporting the robustness of the proposed methodology.

\begin{figure}[h]
    \centering
    \includegraphics[width=\textwidth]{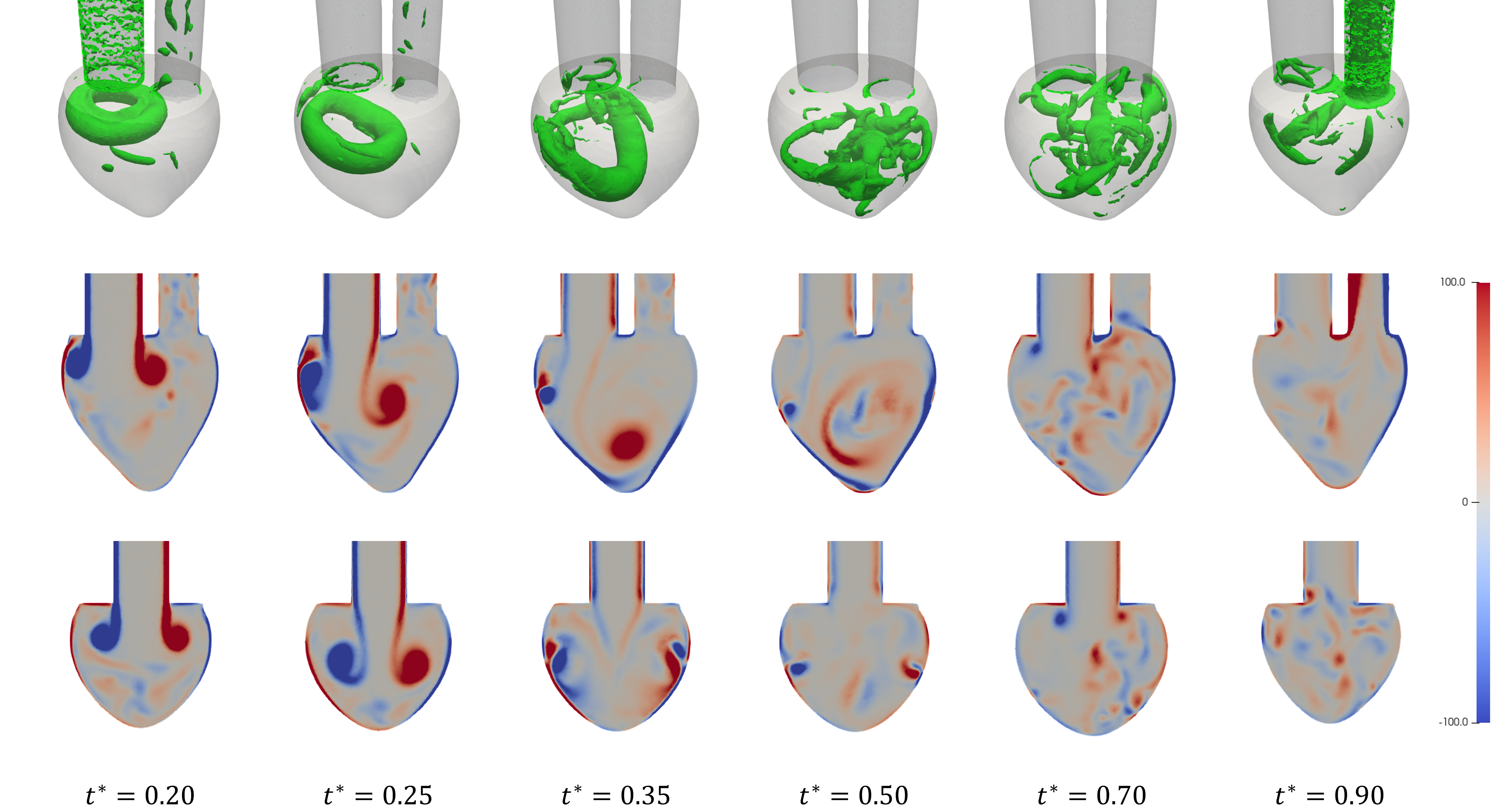}
    \caption{Same as Fig.~\ref{fig:ideal1-flowfeatures} for the \texttt{Ideal 2} database.}
    \label{fig:ideal2-flowfeatures}
\end{figure}

As in \texttt{Ideal 1}, in Fig. \ref{fig:ideal2-flowfeatures} we illustrate the Q-Criterion and plane-normal vorticity to characterize intraventricular flow structures. Top row illustrates the temporal evolution of the vortex ring using a Q-Criterion isocontour at a level of 2000. At $t^* = 0.20$, the VR forms at the mitral valve, initially maintaining a circular shape. The proximity of one side to the ventricular wall causes uneven transport of momentum, leading to an elongation of the vortex ring, as observed in the second frame. The vortex propagates toward the apex, and by $t^* = 0.35$, trailing vortex tubes develop along the wall, inducing instabilities that trigger ring dissipation. A secondary, trailing VR appears due to atrial contraction (A-wave), but it rapidly dissipates because of its low energy and interaction with the residual flow. This behavior is also linked to the fact that the main VR has not fully dissipated, thereby affecting the development of the secondary ring. This contrasts with the previous case, where the main, E-wave induced vortex has practically dissipated when the secondary ring forms during the second inflow. Finally, at $t^* = 0.80$, ventricular contraction begins, the aortic valve opens, and blood is ejected from the left ventricle.

As observed in the evolution of the vortex ring, the breakdown process is slower in this case, allowing the vortex to reach the apex of the LV chamber as it grows and eventually dissipates due to the reduction in velocity and its interaction with the ventricular wall. Compared to the previous model, this breakdown mechanism is notably different, highlighting the significant influence of the ventricular geometry on vortex dynamics. The overall flow behavior also changes, as the vortex ring is able to develop more freely in this configuration, whereas in the previous model it was constrained by the narrower chamber, leading to an earlier and more complex breakup.

A complementary analysis is conducted through plane-normal vorticity ($\omega_y$ for the A-A' plane and $\omega_x$ for the B-B' plane). The middle row presents two-dimensional contours of the $\omega_y$ component in the symmetry plane (A-A'). Initially, VR formation appears symmetric, but as it advances toward the apex, interaction with the ventricular wall induces asymmetry, elongation, and tilting of the structure. The bottom row provides a transverse view of the ventricle, revealing the inherent flow symmetry in this direction.

Comparison of our Q-Criterion and vorticity results with Figures 5, 7, and 8 from  Ref. \cite{Vedula2014Computational} confirms strong agreement with both simulations and experimental observations. These results further validate the accuracy of our simulations and highlight its capability to capture key hemodynamic features in an idealized left ventricle.

\section{\label{sec:Methods} Modal decomposition techniques}

The raw output from the CFD simulations must be preprocessed into a format suitable for our data-driven algorithms. The ventricular geometries considered in this work are more complex and vary over time, thus, we first define a sufficiently large structured grid, essentially a box, that encompasses the maximum volume occupied by the ventricle throughout the cardiac cycle. The simulation data is then interpolated onto this fixed mesh, resulting in a spatiotemporal dataset that includes both fluid-filled and empty regions. This allows for a consistent representation of the data as a five-dimensional tensor of dimensions \( (J_1 \times J_2 \times J_3 \times J_4 \times K) \), where \( J_1 \) corresponds to the number of physical variables (e.g., velocity components, pressure), \( J_2 , J_3 , J_4 \) define the spatial resolution, and \( K \) is the number of temporal snapshots.

For modal decomposition techniques, this tensor is reshaped into a two-dimensional snapshot matrix \( \mathbf{V}\) of size \( J \times K \), where \( J = J_1 \times J_2 \times J_3 \times J_4 \). Each column of this matrix represents the state of the system at a given time, while each row corresponds to a particular spatial location and physical variable. Due to the typically high dimensionality of \( J \) (on the order of \( \mathcal{O}(10^6) \)), this representation is computationally demanding, making dimensionality reduction a crucial step for enabling efficient analysis.

\subsection{\label{subsec:pod} Proper Orthogonal Decomposition}
Proper orthogonal decomposition (POD) \cite{Berkooz1993Proper} is a widely used method for data-driven dimensionality reduction and pattern extraction in complex fluid dynamics problems. In the literature, this term is interchangeable with the singular value decomposition (SVD) \cite{Sirovich1987Turbulence} term since it is one of the two methodologies existent to calculate the POD modes. The method relies on SVD, also known as the snapshot method, to extract the most energetic structures from spatiotemporal data, allowing for an efficient representation of the system with a reduced number of modes.
\newpage
In SVD, we decompose the snapshot matrix and decompose it as follows:
\begin{equation}
    \mathbf{V} = \mathbf{U} \bm{\Sigma} \mathbf{T}^T = \sum_{n=1}^{min(J,K)} \sigma_n \mathbf{u}_n \cdot \mathbf{t}_n^T,
    \label{eq:SVD}
\end{equation}
where \( \mathbf{U} \) contains the spatial modes (POD modes), \( \bm{\Sigma} \) is a diagonal matrix of singular values \( \sigma_1, \dots, \sigma_N \), and \( \mathbf{T} \) is the chronos matrix which contains the temporal coefficients. The symbol \( ()^T \) denotes matrix transposition. A tunable tolerance \( \varepsilon_1 \) determines the number of retained modes \( N \), ensuring that \( \sigma_{N+1}/\sigma_1 \leq \varepsilon_1 \).

The snapshot matrix is then reconstructed using Eq. \ref{eq:SVD} in a reduced dimensionality using the matrices composed of just the truncated modes. This reduction retains only the most energetic structures while filtering out noise and spatial redundancies. The extracted POD modes \( \mathbf{U} \) enable the visualization of the dominant features in the spatiotemporal database, providing insight into the underlying physics. 

When analyzing the singular values obtained from the SVD algorithm, we aim to evaluate the energy contribution of each mode to the overall system dynamics. To this end, we compute the normalized modal energy decay, which assesses how much of the system's total energy is captured by each individual mode. The energy associated with the \(n\)-th mode is given by:

\begin{equation}
    e_n = \frac{\sigma_n^2}{\sum_{n=1}^N \sigma_n^2} \cdot 100 \, .
    \label{eq:energy_svd}
\end{equation}

Moreover, we analyze the temporal evolution of the extracted modes to identify the dominant frequencies present in the system. Given the temporal coefficients associated with each POD mode \( n \), namely \( \textbf{T}_k (t) \), we apply the fast fourier transform (FFT) to decompose these signals into their frequency components, \( f \), obtaining the Fourier transform  \( \mathbf{\hat{T}}_n (t) \). This transformation provides the spectral content of each POD mode, enabling the identification of the fundamental frequencies and their harmonics. The power spectral density (PSD) of each mode can then be computed as $\textbf{P}_n(f) = |\tilde{\textbf{T}}_n(f)|^2$, which quantifies the contribution of different frequency components to each mode. The dominant peaks in the spectra correspond to the most significant oscillatory features in the flow field.

\subsection{\label{subsec:hodmd} Higher Order Dynamic Mode Decomposition}
Dynamic mode decomposition (DMD) \cite{Schmid2010Dynamic} is a powerful tool for extracting temporal patterns in spatiotemporal data. It approximates the system evolution as a linear combination of \( M \) DMD modes, expressed as:
\begin{equation}
    \textbf{v}(x,y,z,t_{k})\simeq  \sum_{m=1}^M a_{m}\textbf{u}_m(x,y,z)e^{(\delta_m+i \omega_m)t_k} \, , \quad \text{for} \; k =1,\ldots ,K \,,
    \label{eq:DMDexpansion}
\end{equation}
where \( \mathbf{u}_m \) are the DMD modes, each with an associated amplitude \( a_m \). These modes evolve according to a growth rate \( \delta_m \) and oscillate at the angular pulsation \( \omega_m \), providing a modal representation of the system's dynamics. As mentioned previously, \(K\) refers to the number of snapshots of the database.

Given a sequence of snapshots \(\{\mathbf{v}_1, \mathbf{v}_2, \dots, \mathbf{v}_K\}\), DMD seeks a best-fit linear operator \(\mathbf{R}\) such that
\begin{equation}
    \mathbf{V}_2^{K} \approx \mathbf{R} \mathbf{V}_1^{K-1} \, .
    \label{eq:Koopman}
\end{equation}

Higher order dynamic mode decomposition (HODMD) \cite{LeClainche2017Higher} addresses several shortcomings of classical DMD regarding robustness and accuracy, particularly for complex fluid flow problems. It incorporates a higher-order approximation of the Koopman operator by linking each snapshot to \( d \) time-delayed snapshots instead of just one. First, we apply SVD to the snapshot matrix \( \mathbf{V}_1^K \), reducing the problem's dimensionality. Then, the reduced temporal matrix is \( \widehat{\mathbf{T}}_1^K = \bm{\Sigma} \textbf{T}^T \), using Eq.~\ref{eq:SVD}. The reduced temporal matrix is then used as input for the DMD-d algorithm, which incorporates a higher-order Koopman assumption, defined as
\begin{equation}
    \widehat{\mathbf{T}}_{d+1}^K \simeq \widehat{\mathbf{R}}_1 \widehat{\mathbf{T}}_1^{K-d}+ \widehat{\mathbf{R}}_2 \widehat{\mathbf{T}}_2^{K-d+1} + \ldots + \widehat{\mathbf{R}}_d \widehat{\mathbf{T}}_d^{K-1}.
    \label{eq:HigherOrderKoopman}
\end{equation}

This formulation generalizes Eq.~\ref{eq:Koopman} by considering delayed correlations across multiple snapshots, enabling HODMD to capture more intricate temporal dynamics than standard DMD. To calibrate the method, we adjust the window parameter \( d \) and the tolerance parameter \( \varepsilon_2 \), which controls the number of modes retained in HODMD via the criterion \( a_{M+1}/a_1 \leq \varepsilon_2 \). It is common practice to use the same tolerance for both POD and HODMD, setting \( \varepsilon_1 = \varepsilon_2 = \varepsilon \), as described in Hetherington \etal~\cite{Hetherington2024Modelflows}.

\section{\label{sec:Results} Results}

In this work, we apply the POD and HODMD modal decomposition techniques described in \S~\ref{sec:Methods} to investigate the underlying physics of the LV simulation datasets. We begin by presenting results for the \texttt{Ideal 1} case, which exhibits more complex dynamics and a faster vortex breakdown process. In contrast, the \texttt{Ideal 2} model features simpler flow dynamics, allowing for an alternative analysis of vortex breakdown.

The analysis of both simulations has been performed on the velocity field. To ensure that only the fully developed flow is analyzed, the initial transient phase is discarded, and the spatiotemporal database is extracted from cycles 10 to 20, where the dynamics have reached a statistically converged state. This choice is supported by previous analyses of the same models, where convergence was verified through modal and spectral consistency across cycles~\cite{Lazpita2025Efficient}. The data were subjected to interpolation, a process which involved the projection of the data onto a structured spatial grid. The dimensions of this grid were \(64 \times 64 \times 128\), with 20 snapshots acquired per cardiac cycle at temporal intervals of \(\Delta t = 0.05\).

As previously mentioned, we employ both POD and HODMD to extract representative modes that reveal key flow features of the vortex ring inside the idealized left ventricle models.

\subsection{\label{subsec:results_pod} Characterization of dominant flow structures via POD}

\begin{figure}[h]
    \centering
    \includegraphics[width=\textwidth]{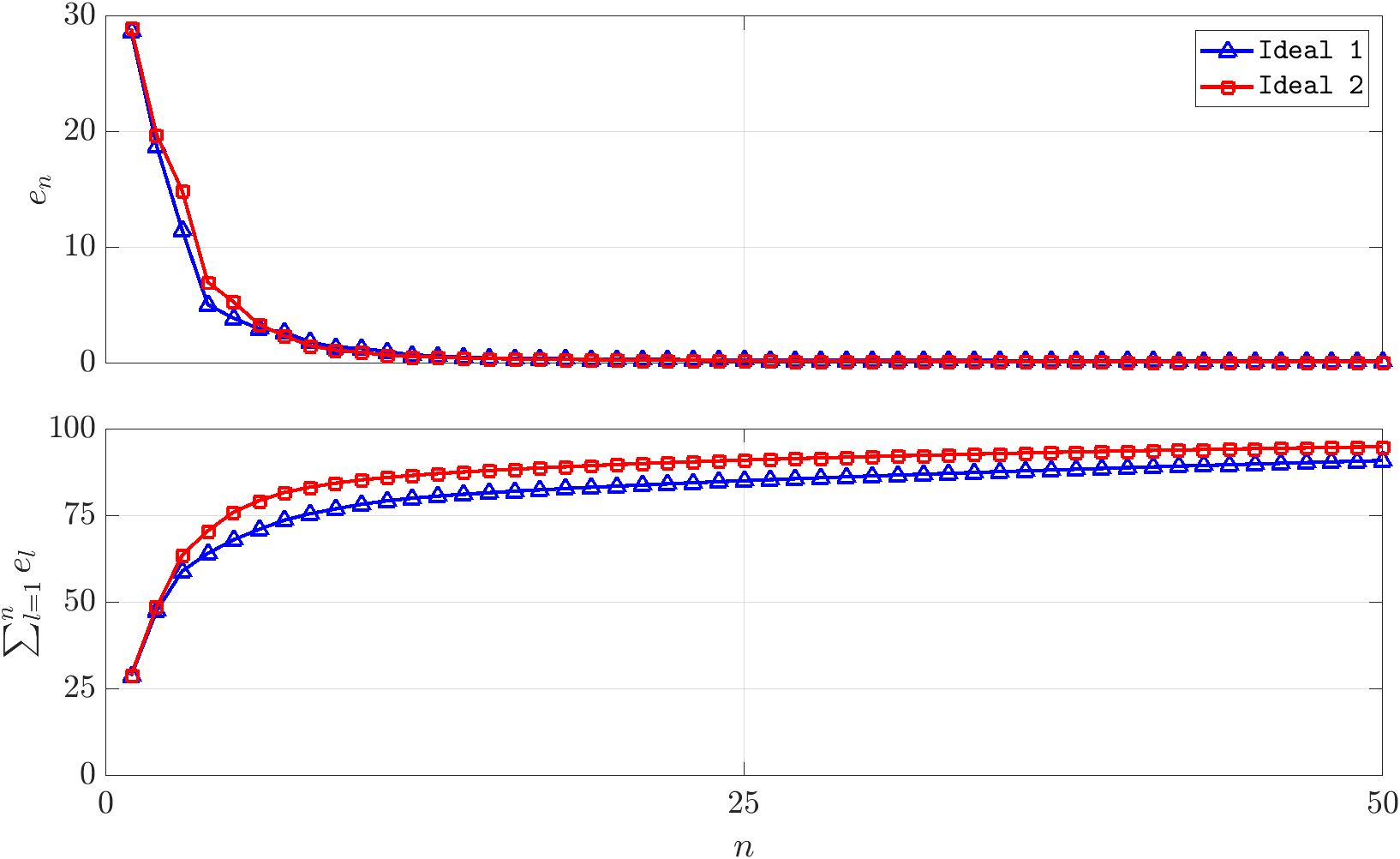}
    \caption{Visualization of energy decay (top) and the cumulative energy (bottom) of the singular values for both idealized left ventricles.}
    \label{fig:svd}
\end{figure}

We begin by applying POD to the simulation data of both idealized models, decomposing the flow fields into a set of orthogonal spatial modes ranked by their associated energy content. Figure~\ref{fig:svd} displays the decay of modal energy (top panel, see Eq.~\ref{eq:energy_svd}) and the corresponding cumulative energy (bottom panel) for both left ventricle geometries.

In the top panel, a steep initial decay in the singular values is observed, indicating that most of the flow's dynamic content is concentrated in the first 15 modes. The energy decay is comparable across both models, although the second geometry exhibits slightly higher energy concentration in the first five modes. In contrast to the observations typically made for the most elementary periodic flows (see Ref.~\cite{Begiashvili2023Data}), in this case there is an absence of singular values organized in energetically paired modes. This suggests a higher level of dynamical complexity in the intraventricular flow, as will be further discussed in subsequent sections.

Turning to the bottom panel, which presents the cumulative energy distribution, the dominance of the leading modes becomes even more evident. In both cases, the first 25 modes suffice to capture nearly 95\% of the total energy. For clarity, only the first 50 modes are shown, as the contribution of higher-order modes is negligible.

Given that the period of our simulation is defined as \( T = 2\pi/f \), all primary flow features (\ie the formation of the main vortex ring, the tilting of such vortex and its eventual dissipation, the appearance of secondary vortical structures, and the final outflow) occur at the fundamental frequency \( f = 2\pi \; \text{Hz}\). However, the cardiac cycle exhibits two distinct inflow phases, corresponding to the E and A-waves, which take place approximately every \( T/2 \). This periodicity introduces a secondary relevant frequency in the system, namely \( f = 4\pi \), associated with the double-pulsed inflow pattern.

\begin{figure}[h]
    \centering
    \includegraphics[width=\textwidth]{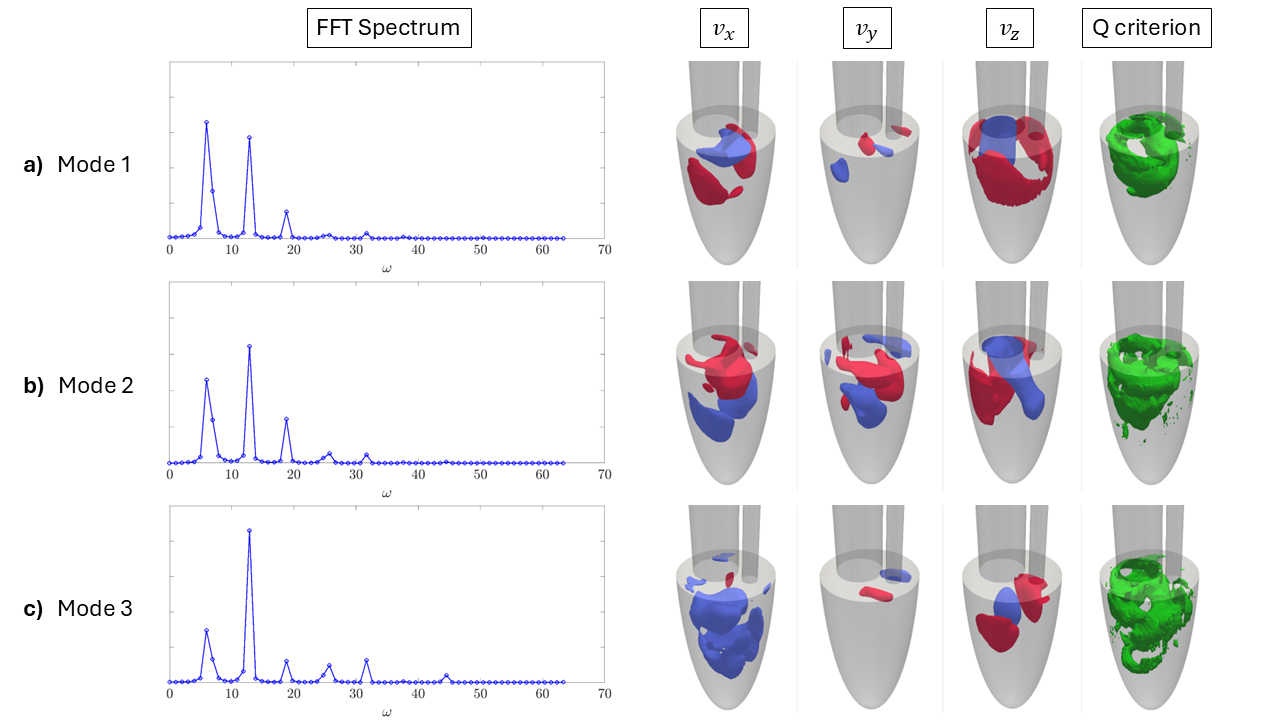}
    \caption{From left to right: FFT spectra, three-dimensional isocontours of the normalized POD modes, showing velocity components \( v_x, v_y \text{ and } v_z\) (-0.6 blue, 0.6 red) and Q-criterion (2000 green) for the first three most energetic modes of the \texttt{Ideal 1} model.}
    \label{fig:ideal1-pod3d-1}
\end{figure}

The results are summarized in Fig. \ref{fig:ideal1-pod3d-1}, which presents FFT spectra of the chronos matrix \( \mathbf{T}\) and three-dimensional isocontours of the velocity field for the three most energetic spatial support of the modes, alongside Q-criterion isosurfaces extracted from these modes to improve the visualization of the vortical features.

Examining the FFT spectra (Fig.~\ref{fig:ideal1-pod3d-1}, left), we observe distinct peaks at \( 2\pi \) and its harmonics (\( 4\pi \), \( 6\pi \), and so forth), which is consistent with the periodic nature of the simulation. As discussed previously, the peak at \( 2\pi \) corresponds to the fundamental frequency of the cardiac cycle, while the presence of a significant peak at \( 4\pi \) may be attributed to the double-pulsed inflow dynamics, specifically the E-wave and A-wave, occurring within each cycle.

Analyzing the modes individually, the most energetic mode exhibits dominant contributions at the fundamental frequency \( 2\pi \), along with a secondary peak at \( 4\pi \), and a smaller contribution at \( 6\pi \). This suggests that the mode primarily captures the periodicity of the VR dynamics while also incorporating double-pulse and other non-linear effects. The corresponding 3D isocontours of the velocity confirm this, showing strong recirculation near the vortex ring. The Q-criterion isosurface pinpoints the early tilting phase of the ring.

The second mode, as presented in Fig.~\ref{fig:ideal1-pod3d-1}(b), shares similarities with the first but now the amplitude referred to the \( 2\pi \) frequency has decreased and the most pronounced one is the $4\pi$ frequency. The \( 6\pi \) frequency maintains a small contribution. The change in the most relevant frequency is evident in its velocity field, which extends beyond the vortex formation region. The Q-criterion visualization reveals both the formation and progression of the ring along the geometry.

The third mode is predominantly characterized by a \( 4\pi \) frequency; however, a moderate contribution from the fundamental frequency \( 2\pi \) but also smaller contributions from its higher harmonics (\( 6\pi \), \( 8\pi \), and \( 10\pi \)) are also observed. This predominance of the \( 4\pi \) frequency is reflected in the Q-criterion isosurface, which reveals coherent vortical structures near both the inlet and downstream regions; these structures stretch towards the ventricular apex. Furthermore, the vortex ring is observed to advance through the geometry. The influence of higher harmonics must be considered, as vortex breakdown depends on nonlinear interactions between frequencies, among other phenomena \cite{Lazpita2022Generation}.

As demonstrated in Fig.~\ref{fig:ideal1-flowfeatures}, the geometry itself impedes the complete development of the vortex ring, thereby initiating an early breakdown. This, in turn, leads to an escalation in flow complexity and an enhancement in nonlinear interaction among modes.

\begin{figure}[h]
    \centering
    \includegraphics[width=\textwidth]{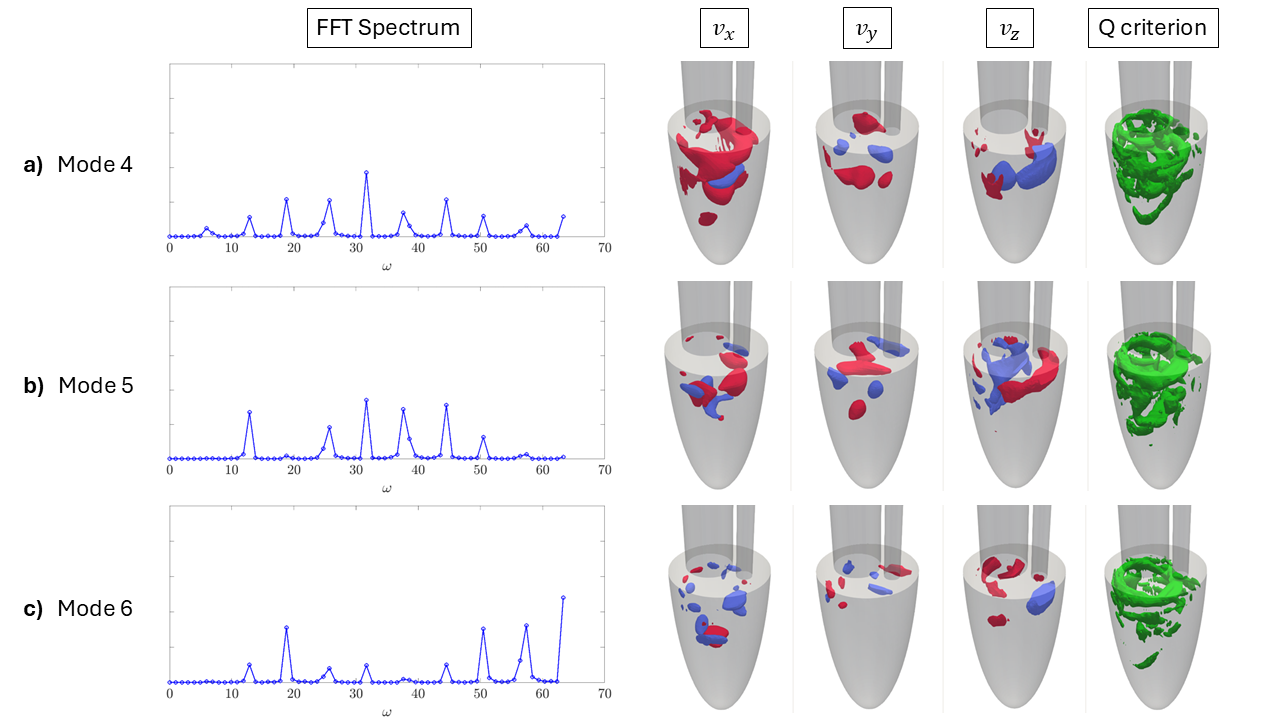}
    \caption{Same as Fig. \ref{fig:ideal1-pod3d-1} for the 4 to 6 most energetic modes.}
    \label{fig:ideal1-pod3d-2}
\end{figure}

Figure \ref{fig:ideal1-pod3d-2} illustrates how modes 4 to 6 capture the loss of coherence of the primary vortex and the early dissipation of the secondary vortex, highlighting the emergence of higher harmonics associated with smaller spatial scales and higher frequencies. Although these modes carry less energy, they provide insight into the breakdown and dissipation mechanisms of the main flow structures.

\begin{figure}[h]
    \centering
    \includegraphics[width=\textwidth]{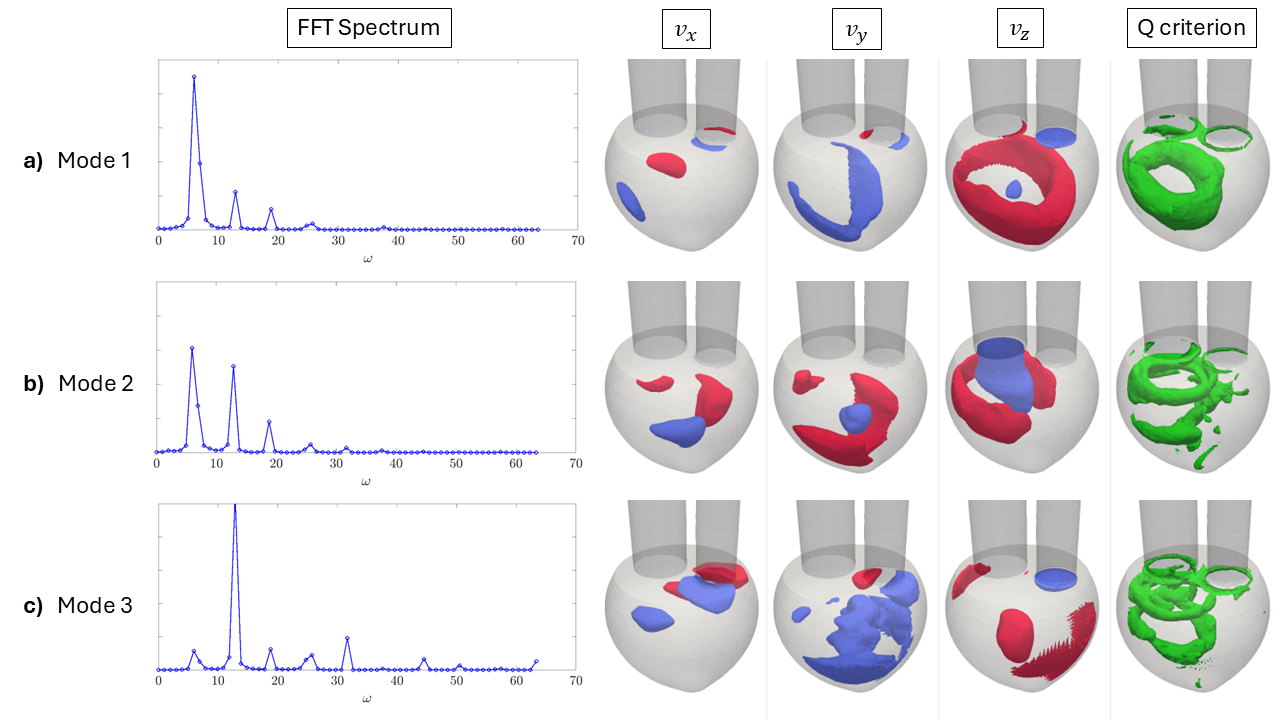}
    \caption{Counterpart of Fig. \ref{fig:ideal1-pod3d-1} for the \texttt{Ideal 2} database.}
    \label{fig:ideal2-pod3d-1}
\end{figure}

Next, the same analysis of the POD spatial mode support and the FFT analysis of the chronos matrix is applied to the second ideal case, which reveals some differences compared to the previous case. Figure \ref{fig:ideal2-pod3d-1} presents the FFT spectra along with three-dimensional isocontours of the velocity field for the three most energetic modes. Additionally, Q-criterion isosurfaces have been extracted from these modes to enhance the visualization of vortical structures. Similarly, as in the previous case, the dominant frequency (\(2\pi\)) associated with the heart rate periodicity is clearly captured, along with the \(4\pi\) frequency related to the double inflow pulse, and their higher harmonics. However, the dynamics in this case are significantly less entangled. The first mode predominantly represents the dominant frequency, while the third mode is distinctly associated with the \( 4\pi \) frequency. In contrast, the second mode shows contributions from both frequencies in equal measure. Lower-energy modes exhibit nonlinear interactions between different frequencies, though they primarily involve lower harmonics rather than higher ones, as observed in Fig. \ref{fig:ideal1-pod3d-2}. To capture higher frequency interactions, it would be necessary to study modes with less energetic contribution than the ones presented.

These findings are further supported by the 3D isocontour representations of velocity and Q-criterion fields associated with each POD mode. The first mode, which captures the fundamental periodicity at \(2\pi\), illustrates the recirculating flow around the main vortical ring, clearly visible in the Q-criterion plots. This structure extends along the ventricle, consistent with Fig.~\ref{fig:ideal2-flowfeatures}, where the vortex ring was shown to travel toward the apex before breaking down.

The second mode reflects an interaction between \(2\pi\) and \(4\pi\) frequencies, revealing the formation of a vortex at the inlet and the dissipation of the main structure downstream (towards the apex). The third mode is predominantly associated with the \(4\pi\) frequency and likely represents the double-pulsing behavior observed in the simulations. Its Q-criterion isosurface shows a VR near the apex and another forming at the inflow region, suggesting a connection with the secondary filling phase.

Unlike the previous case, where the spatial support of the modes was mostly concentrated near the inlet (see Fig.~\ref{fig:ideal1-pod3d-1}), the modes in this geometry span along the entire ventricular domain. This reflects a more complete development of the VR: the ring progesses relatively undisturbed for a greater length through the ventricle before undergoing breakdown. Note how modes 1–3 appear primarily related to the formation and evolution of the main vortex ring and its interaction with the secondary vortex.

\begin{figure}[h]
    \centering
    \includegraphics[width=\textwidth]{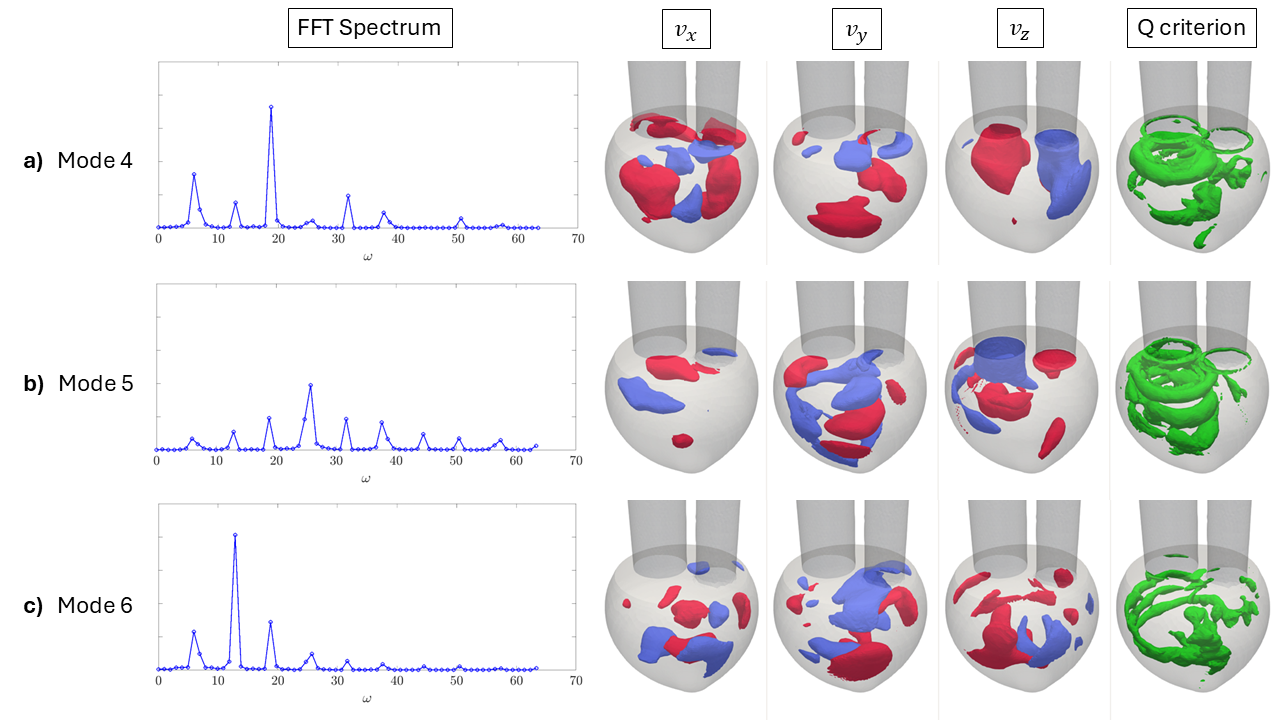}
    \caption{Counterpart of Fig. \ref{fig:ideal1-pod3d-2} for the \texttt{Ideal 2} database.}
    \label{fig:ideal2-pod3d-2}
\end{figure}

In contrast, modes 4 and 5 shown in Fig.~\ref{fig:ideal1-pod3d-2}, characterized by higher harmonic content, capture finer structures and showcase larger spatial support, suggesting they arise from nonlinear interactions that contribute to vortex breakdown. Mode 6 again highlights the presence of the secondary vortex, with a subtle influence from the primary structure, though with reduced energy. Interestingly, this mode resembles mode 3 in structure but appears at a lower energy level, possibly indicating a redistribution of energy among similar flow features across different frequencies.

\subsection{\label{subsec:results_hodmd} Identification of dynamical features via HODMD}

\begin{figure}[h]
    \centering
    \includegraphics[width=\textwidth]{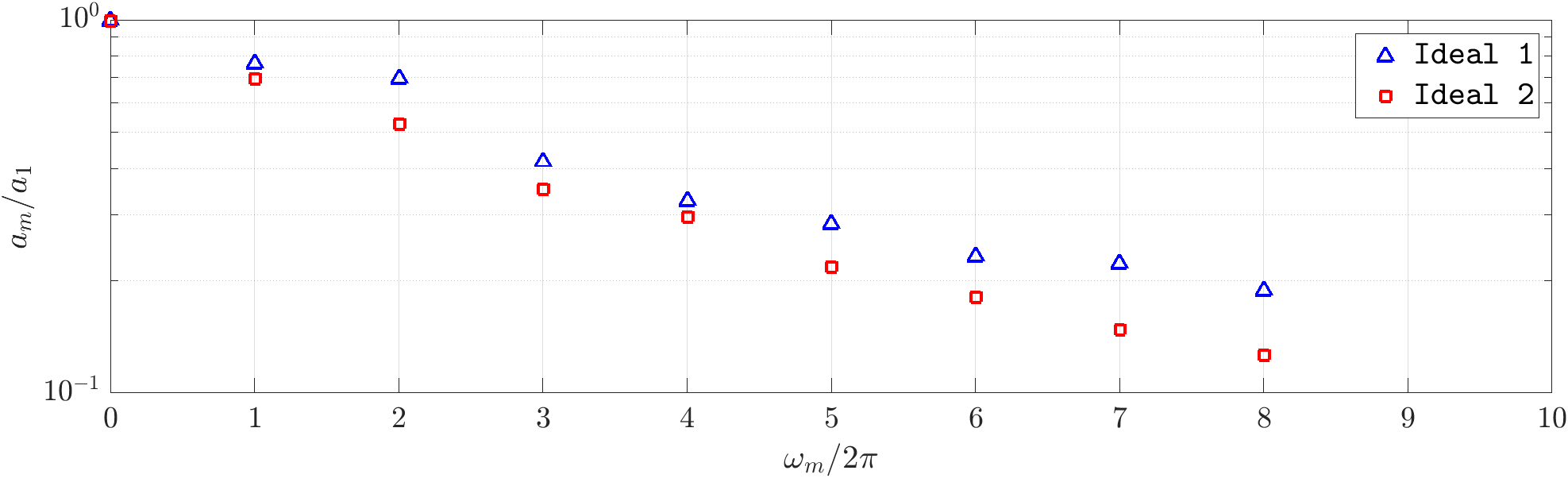}
    \caption{Normalized amplitude spectra of the HODMD modes as a function of the angular pulsation divided by the main frequency (\(2\pi\)) for both idealized left ventricle models. Only robust modes, identified through calibration of the window size \( d \) and tolerance \( \varepsilon \), are shown.}
    \label{fig:hodmd-spectrum}
\end{figure}

As previously discussed, the dataset displays a complex spectral structure. To enhance the frequency analysis beyond what is captured by POD and FFT, we apply the HODMD, introduced in Sec.~\ref{subsec:hodmd}. This method represents the data as a linear combination of modes, each associated with a distinct temporal frequency. Figure~\ref{fig:hodmd-spectrum} shows the normalized amplitude spectrum of the resulting DMD modes as a function of frequency, based on a representative calibration with carefully selected values of the tolerance \( \varepsilon \) and window size \( d \). These parameters were chosen through an exhaustive exploration of the calibration space (\(d\)--\( \varepsilon\)), following the protocol outlined in Ref.~\cite{LeClainche2017Higher}.

To facilitate comparison, only the robust and physically relevant modes are shown, namely, the dominant frequencies and their harmonics, which appear as well-defined clusters of high amplitude. Spurious modes and numerical noise, which do not correspond to meaningful flow features, have been filtered out to highlight the essential dynamics. Frequencies have been non-dimensionalized by dividing them by the main frequency (\(2\pi\)), thereby facilitating the more precise identification of harmonic modes.

A particularly interesting observation is made in the \texttt{Ideal 1} geometry: the amplitudes at the fundamental frequency (\(2\pi\)) and its first harmonic (\(4\pi\)) are of similar magnitude, while higher harmonics decay more gradually. This suggests that the occurrence of a secondary inflow event, linked to the formation of a second vortex ring, is dynamically as relevant as the overall periodicity of the cardiac cycle. In contrast, the \texttt{Ideal 2} model exhibits a more uniform decay across all harmonics, with the dominant mode remaining clearly predominant. Moreover, the decay is steeper in this case, as all amplitudes remain lower than those in \texttt{Ideal 1}, indicating a reduced level of temporal complexity.

To visualize the DMD modes, it is first necessary to introduce the concept of mode phase. Instead of separately displaying the real and imaginary parts of the velocity components, we reconstruct the velocity field at various phase instants and visualize the corresponding Q-criterion isocontours. The phase is defined as the product of the angular frequency and time, \( \phi = \omega_m t_k \), where \( \phi \in [0,2\pi) \). This representation allows us to observe the evolution in time of coherent structures across all DMD modes. The phase consists of \( 2\pi \) where \( \phi = 0 = 2\pi \) corresponds to the real part of the mode, \( \phi = \pi/2 \) to the positive imaginary part, and at \( \phi = 3\pi/2 \) we have the negative imaginary part.

\begin{figure}[h]
    \centering
    \includegraphics[trim=0 0 350 0, clip, width=0.8\textwidth]{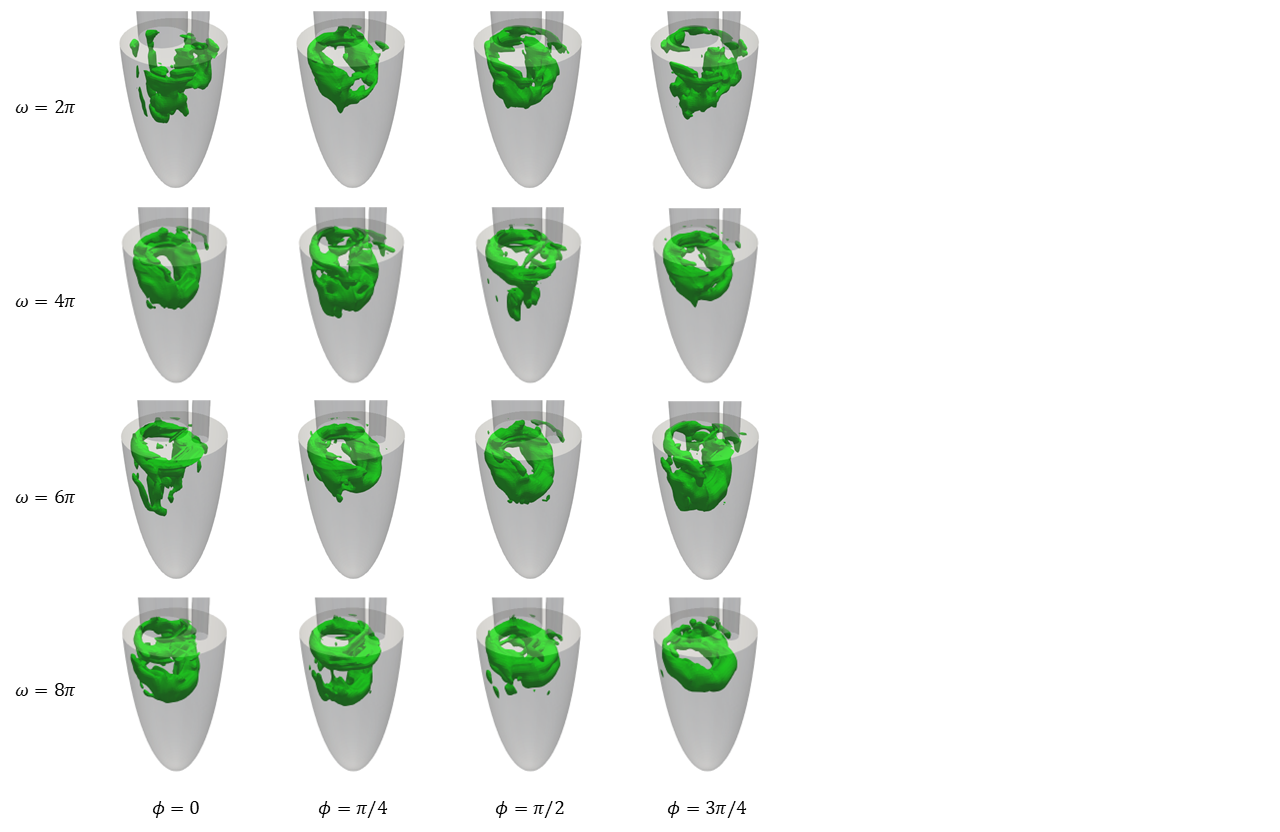}
    \caption{Three-dimensional Q-criterion isocontours at level 2000 for the dominant DMD mode and its first three harmonics in each row identified using their associated frequencies \( \omega \), reconstructed at different phase instants (\( \phi \)), in the \texttt{Ideal 1} model.}
    \label{fig:ideal1-dmd-phase1}
\end{figure}

The reconstructed structures for the largest amplitude mode and its first three harmonics are shown in Fig.~\ref{fig:ideal1-dmd-phase1}, at four different phase instants. Focusing on the dominant mode, we observe the evolution of a single vortical ring. In contrast, the mode with frequency \( \omega = 4\pi \) reveals the generation of a second ring interacting with the main vortex ring at \( \phi = \pi/2 \), supporting the hypothesis that this frequency is associated with the double-pulsing behavior. This observation is further reinforced by the next two harmonics. The structure at \( \omega = 6\pi \), a harmonic of the dominant frequency but not of the double-pulsing mode, still exhibits a single VR, although traces of the second ring can also be observed. Finally, in the mode with frequency \( \omega = 8\pi \), a harmonic of the double-pulsing frequency, clearly exhibits the formation of two distinct vortex rings within the ventricle.

\begin{figure}[h]
    \centering
    \includegraphics[trim=0 0 350 0, clip, width=0.8\textwidth]{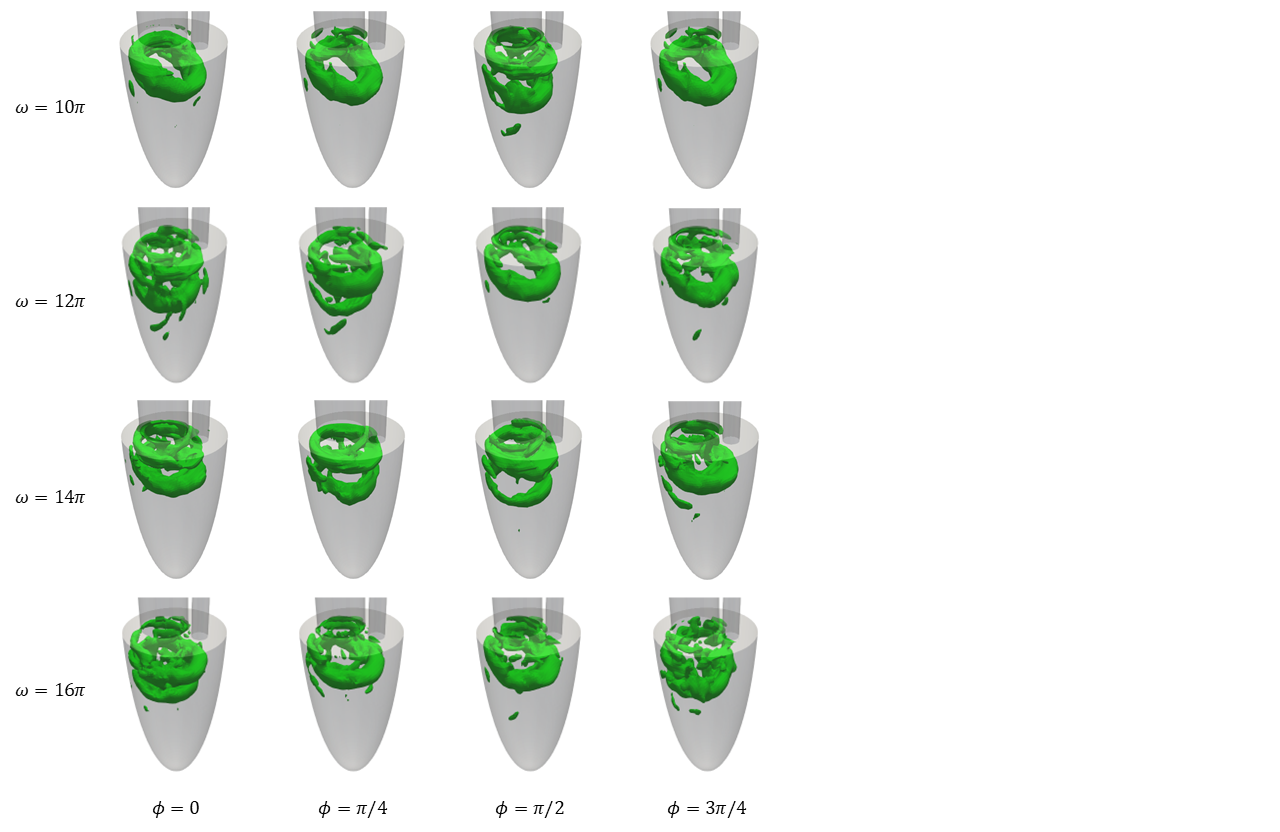}
    \caption{Same as Fig. \ref{fig:ideal1-dmd-phase1} for the higher harmonics.}
    \label{fig:ideal1-dmd-phase2}
\end{figure}

Figure~\ref{fig:ideal1-dmd-phase2} shows the same representation for the next four higher frequency harmonics. These modes exhibit more fragmented and spatially intricate structures, indicating finer scales. Nonetheless, the double-pulsing characteristics remain visible, which may be driven by nonlinear interactions between the dominant mode and its first harmonic.

\begin{figure}[h]
    \centering
    \includegraphics[trim=0 0 300 0, clip, width=0.8\textwidth]{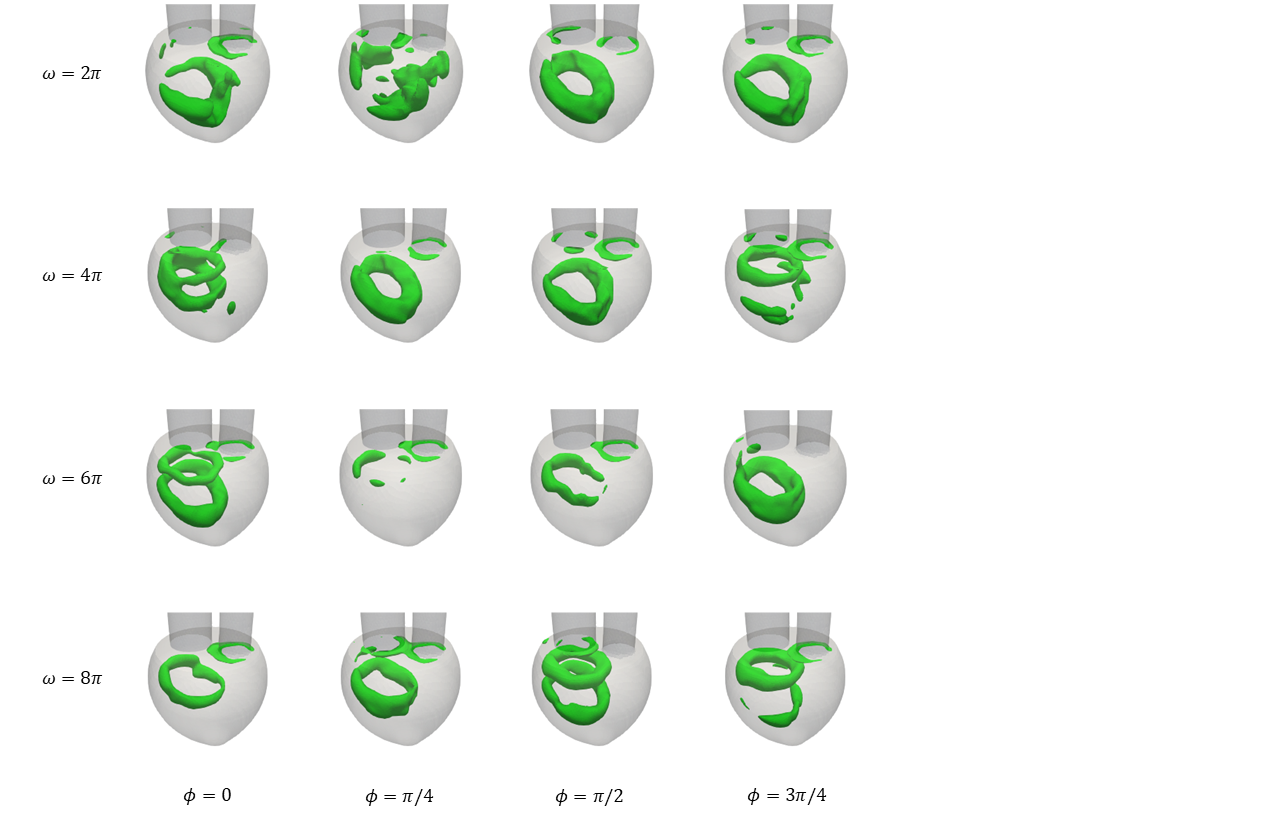}
    \caption{Three-dimensional Q-criterion isocontours at level 2000 for the dominant DMD mode and its first three harmonics in each row identified using their associated frequencies \( \omega \), reconstructed at different phase instants (\( \phi \)), in the \texttt{Ideal 2} model.}
    \label{fig:ideal2-dmd-phase1}
\end{figure}

Similar to the \texttt{Ideal 1} case, the DMD analysis for the \texttt{Ideal 2} model is based on visualizing the Q-criterion isocontours of each mode at different phase instants, as shown in Fig.~\ref{fig:ideal2-dmd-phase1}. At first glance, the flow features appear cleaner and more coherent compared to the previous case, which can be attributed to the slower and more regular dynamics of the simulation.

As before, the dominant mode reveals the presence of the primary vortex ring, consistent with the periodicity of the flow. For $\omega=4\pi$, in this case, two distinct vortex rings are already visible at the initial phase instant, indicating a clearer presence of the double-pulsing phenomenon. The nonlinear interaction between the first two frequencies is more pronounced than in \texttt{Ideal 1}, particularly evident in the mode at \( \omega = 6\pi \), where two rings are again observed, one near the inlet and another toward the center of the ventricle.

The double-pulse phenomena becomes even more evident in \( 8\pi \). As the phase evolves, the first vortex ring moves downward while a second ring begins to form at the inlet. Once fully formed, the first ring starts to dissipate, clearly showing a cyclic generation-dissipation process consistent with a double-pulsating pattern.

\begin{figure}[h]
    \centering
    \includegraphics[trim=0 0 300 0, clip, width=0.8\textwidth]{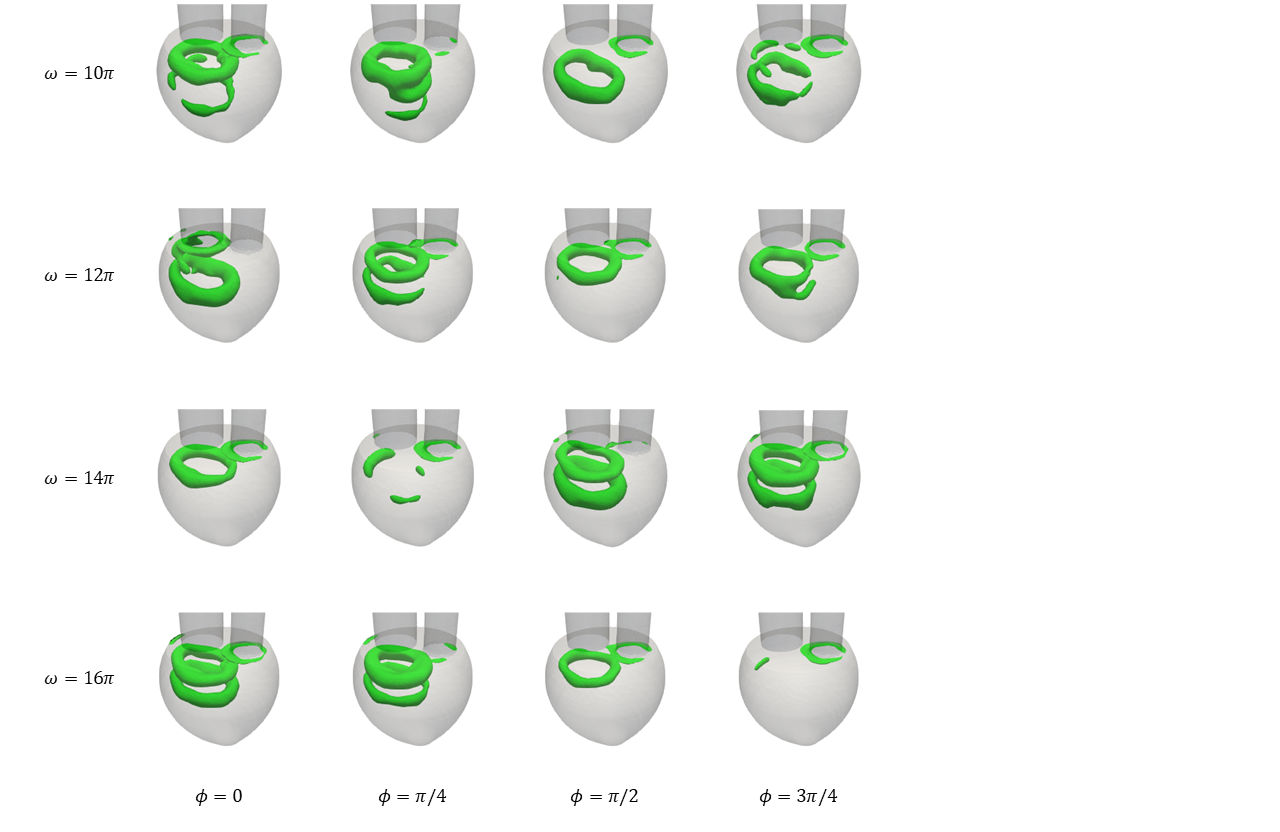}
    \caption{Same as Fig. \ref{fig:ideal2-dmd-phase1} for the higher harmonics.}
    \label{fig:ideal2-dmd-phase2}
\end{figure}

Figure~\ref{fig:ideal2-dmd-phase2} displays the higher-order harmonics. The emergence of modes whose spatial support reflects features from both the dominant periodic behavior and the double-pulsating component supports the idea of a strong interaction between these two mechanisms. This is further illustrated by the three-dimensional Q-criterion isocontours, which frequently reveal the coexistence of two vortex rings within the left ventricle, along with clear evidence of vortex generation, dissipation, and interaction near the outlet region.

Finally, a comparison between the HODMD and FFT analysis of the POD chronos modes reveals a strong spectral correspondence. For the \texttt{Ideal 1} case, the DMD modes with frequencies ranging from \( \omega = 2\pi \) to \( \omega = 8\pi \) are primarily captured by the first three POD modes, which represent the main vortex ring, its initial breakdown, and the onset of the double-pulse. In contrast, higher-frequency DMD modes (e.g., \( \omega = 10\pi \) and beyond) are more closely related to POD modes 4 to 6, associated with smaller-scale structures and later stages of the flow evolution. This suggests again that the interaction of the secondary vortex with the main vortex ring is one of the effects influencing the main vortex breakdown.

For the \texttt{Ideal 2} model, the first POD mode captures the main vortex ring, including its characteristic tilting and recirculation patterns, and aligns well with the dominant DMD mode associated with the fundamental frequency at \(2\pi\). The third POD mode, which presents two spatially distinct vortical structures, is consistent with the double-pulsing behavior observed in the simulations. The second POD mode appears as a nonlinear combination of both the main and secondary vortex contributions. Lower-energy POD modes exhibit smaller-scale structures, indicative of nonlinear interactions.

Note that the flow behaviors discussed above, such as the interaction between dominant and secondary components, were already suggested by the earlier POD analysis, which revealed spatially dominant patterns. However, the DMD framework enables a more explicit characterization of their temporal evolution, as each mode is associated with a distinct angular frequency and a corresponding growth or decay rate. This allows for a clearer interpretation of the dynamics: the primary vortex ring is captured by the mode at \(2\pi\), while the emergence of a secondary structure becomes apparent near \(4\pi\), eventually becoming dominant at \(8\pi\). From that point onward, both vortices are consistently present across higher-frequency modes, confirming their coexistence and sustained interaction throughout the cardiac cycle.

\section{\label{sec:Conclusions} Conclusions}

This study investigates the fluid dynamics underlying vortex ring formation and breakdown in the left ventricle (LV) during early diastole, using computational fluid dynamics (CFD) simulations of two idealized geometries under physiological inflow conditions. To the authors’ knowledge, this is the first time such an analysis has been conducted with the explicit goal of understanding the physics of vortex ring evolution in idealized ventricular chambers. The comparative approach between two different geometric configurations reveals how key flow mechanisms, particularly the formation, propagation, and breakdown of the vortex ring, are heavily influenced by the geometry itself.

In the \texttt{Ideal 1} model, which features a narrow geometry and a larger stroke volume, the stronger pressure gradient leads to higher inflow velocities. This promotes the formation of a more intense vortex ring that interacts early with the chamber walls. As a result, vortex breakdown occurs prematurely and violently, giving rise to a more complex flow regime with stronger nonlinear interactions. In contrast, the \texttt{Ideal 2} model consists of a comparatively wider cavity and a relatively smaller volume change. These features allow the primary vortex ring to travel farther, reaching the apex before gradually breaking down due to the reduction in inflow momentum, rather than wall interaction. This leads to a smoother and more coherent flow evolution.

To further analyze these dynamics, two modal decomposition techniques, proper orthogonal decomposition (POD) and higher order dynamic mode decomposition (HODMD), were employed. These tools enable a reduced-order description of the flow while retaining key physical information, helping to identify the dominant structures and their temporal behavior. The analysis reveals two main frequencies across both cases: a natural frequency at $\omega = 2\pi$ connected to the heart beat rate, linked to the generation and evolution of the main vortex ring, and a harmonic at $\omega = 4\pi$, associated with the secondary inflow pulse that gives rise to a second, trailing vortex ring. Higher harmonics were also identified: they carry less energy and manifest primarily through nonlinear interactions in the flow.

In \texttt{Ideal 1} case, the early breakdown of the main vortex ring increases the complexity of the flow, leading to stronger interactions between frequency components. As a result, traces of higher harmonics are observed to corrupt even the most energetic POD modes, and spatial structures appear more fragmented. In contrast, \texttt{Ideal 2} displays cleaner, more easily interpretable modes, with the principal POD modes more clearly associated with the dominant temporal frequencies and less influenced by higher-order components.

The HODMD analysis, by providing a factorization of the data in temporal components distinctly associated with an angular pulsation \( \omega_m \) and growth rate \( \delta_m \),  offers a complementary view of the flow. In both geometries, this algorithm clearly identifies modes associated with the presence and motion of both vortex rings. In \texttt{Ideal 2}, these modes are cleanly separated, each capturing a different stage of vortex evolution, from formation to dissipation. In \texttt{Ideal 1}, however, the modal content reflects stronger coupling and interference between frequencies, consistent with the more abrupt and chaotic vortex dynamics observed.

These findings highlight the critical role of geometry in shaping intraventricular flow dynamics. Depending on the chamber shape and volume variation, the dominant mechanisms driving vortex formation and decay can vary significantly. This insight can guide researchers and clinicians in selecting the most appropriate idealized geometry for their studies and help interpret patient-specific data when coupled with high-fidelity CFD simulations.

Future work will focus on extending this framework to patient-specific geometries, including pathological cases. The present study provides a foundational reference for identifying deviations from healthy flow conditions and offers a valuable toolset for characterizing hemodynamic patterns in the human heart.

\section{Data Availability Statement}
The data that support the findings of this study are available upon reasonable request.
\\
Should the reader wish to gain a more detailed understanding of the process by which the databases were obtained, as well as other pertinent information, they are invited to visit the following website, where they will find the relevant codes and tutorials in Ref. \cite{ModelFLOWsCardiac}.

\section{Acknowledgements}
The authors acknowledge the grant PLEC2022-009235 funded by MCIN/AEI/ 10.13039/501100011033 and by the European Union “NextGenerationEU”/PRTR and the grant PID2023-147790OB-I00 funded by MCIU/AEI/10.13039/501100011033/FEDER, UE. The authors gratefully acknowledge the Universidad Politécnica de Madrid (www.upm.es) for providing computing resources on Magerit Supercomputer.

The authors gratefully acknowledge Prof. Vedula for kindly providing the geometry data of \texttt{Ideal 2} model that was essential for conducting this study.

\bibliography{biblio}

\begin{thebibliography}{10}
\expandafter\ifx\csname url\endcsname\relax
  \def\url#1{\texttt{#1}}\fi
\expandafter\ifx\csname urlprefix\endcsname\relax\def\urlprefix{URL }\fi
\expandafter\ifx\csname href\endcsname\relax
  \def\href#1#2{#2} \def\path#1{#1}\fi

\bibitem{WHO2024}
{World Health Organization}, Cardiovascular diseases (cvds) - fact sheet,
  \url{https://www.who.int/news-room/fact-sheets/detail/cardiovascular-diseases-(cvds)}
  (2024).

\bibitem{Pedrizzetti2014Vortex}
G.~Pedrizzetti, G.~La~Canna, O.~Alfieri, G.~Tonti,
  \href{https://doi.org/10.1038/nrcardio.2014.75}{The vortex—an early
  predictor of cardiovascular outcome?}, Nature Reviews Cardiology 11~(9)
  (2014) 545--553.
\newblock \href {https://doi.org/10.1038/nrcardio.2014.75}
  {\path{doi:10.1038/nrcardio.2014.75}}.
\newline\urlprefix\url{https://doi.org/10.1038/nrcardio.2014.75}

\bibitem{Toger2012Vortex}
J.~Töger, M.~Kanski, M.~Carlsson, S.~J. Kovács, G.~Söderlind, H.~Arheden,
  E.~Heiberg, \href{https://doi.org/10.1007/s10439-012-0615-3}{Vortex ring
  formation in the left ventricle of the heart: analysis by 4d flow mri and
  lagrangian coherent structures}, Annals of Biomedical Engineering 40 (2012)
  2652--2662.
\newblock \href {https://doi.org/10.1007/s10439-012-0615-3}
  {\path{doi:10.1007/s10439-012-0615-3}}.
\newline\urlprefix\url{https://doi.org/10.1007/s10439-012-0615-3}

\bibitem{Le2012Vortex}
T.~Le, F.~Sotiropoulos, D.~Coffey, D.~Keefe,
  \href{https://doi.org/10.1063/1.4747164}{Vortex formation and instability in
  the left ventricle}, Physics of Fluids 24~(9) (2012).
\newblock \href {https://doi.org/10.1063/1.4747164}
  {\path{doi:10.1063/1.4747164}}.
\newline\urlprefix\url{https://doi.org/10.1063/1.4747164}

\bibitem{Grunwald2022Intraventricular}
A.~Grünwald, J.~Korte, N.~Wilmanns, C.~Winkler, K.~Linden, U.~Herberg,
  S.~Groß-Hardt, U.~Steinseifer, M.~Neidlin,
  \href{https://doi.org/10.1007/s13239-021-00598-9}{Intraventricular flow
  simulations in singular right ventricles reveal deteriorated washout and low
  vortex formation}, Cardiovascular Engineering and Technology (2022) 1--9\href
  {https://doi.org/10.1007/s13239-021-00598-9}
  {\path{doi:10.1007/s13239-021-00598-9}}.
\newline\urlprefix\url{https://doi.org/10.1007/s13239-021-00598-9}

\bibitem{Korte2023Hemodynamic}
J.~Korte, T.~Rauwolf, J.~Thiel, A.~Mitrasch, P.~Groschopp, M.~Neidlin,
  A.~Schmeißer, R.~Braun-Dullaeus, P.~Berg,
  \href{https://doi.org/10.3390/fluids8020071}{Hemodynamic assessment of the
  pathological left ventricle function under rest and exercise conditions},
  Fluids 8~(2) (2023) 71.
\newblock \href {https://doi.org/10.3390/fluids8020071}
  {\path{doi:10.3390/fluids8020071}}.
\newline\urlprefix\url{https://doi.org/10.3390/fluids8020071}

\bibitem{Fedele2023Comprehensive}
M.~Fedele, R.~Piersanti, F.~Regazzoni, M.~Salvador, P.~Africa, M.~Bucelli,
  A.~Zingaro, A.~Quarteroni, \href{https://doi.org/10.1016/j.cma.2023.115983}{A
  comprehensive and biophysically detailed computational model of the whole
  human heart electromechanics}, Computer Methods in Applied Mechanics and
  Engineering 410 (2023) 115983.
\newblock \href {https://doi.org/10.1016/j.cma.2023.115983}
  {\path{doi:10.1016/j.cma.2023.115983}}.
\newline\urlprefix\url{https://doi.org/10.1016/j.cma.2023.115983}

\bibitem{Zingaro2024Electromechanics}
A.~Zingaro, M.~Bucelli, R.~Piersanti, F.~Regazzoni, A.~Quarteroni,
  \href{https://doi.org/10.1016/j.jcp.2024.112885}{An electromechanics-driven
  fluid dynamics model for the simulation of the whole human heart}, Journal of
  Computational Physics 504 (2024) 112885.
\newblock \href {https://doi.org/10.1016/j.jcp.2024.112885}
  {\path{doi:10.1016/j.jcp.2024.112885}}.
\newline\urlprefix\url{https://doi.org/10.1016/j.jcp.2024.112885}

\bibitem{Schenkel2009MRI}
T.~Schenkel, M.~Malve, M.~Reik, M.~Markl, B.~Jung, H.~Oertel,
  \href{https://doi.org/10.1007/s10439-008-9627-4}{Mri-based cfd analysis of
  flow in a human left ventricle: Methodology and application to a healthy
  heart}, Annals of Biomedical Engineering 37 (2009) 503--515.
\newblock \href {https://doi.org/10.1007/s10439-008-9627-4}
  {\path{doi:10.1007/s10439-008-9627-4}}.
\newline\urlprefix\url{https://doi.org/10.1007/s10439-008-9627-4}

\bibitem{Nguyen2015Patient}
V.~T. Nguyen, S.~N. Wibowo, Y.~A. Leow, H.~H. Nguyen, Z.~Liang, H.~L. Leo,
  \href{https://doi.org/10.1007/s13239-015-0244-8}{A patient-specific
  computational fluid dynamic model for hemodynamic analysis of left ventricle
  diastolic dysfunctions}, Cardiovascular Engineering and Technology 6 (2015)
  412--429.
\newblock \href {https://doi.org/10.1007/s13239-015-0244-8}
  {\path{doi:10.1007/s13239-015-0244-8}}.
\newline\urlprefix\url{https://doi.org/10.1007/s13239-015-0244-8}

\bibitem{Colorado2022Patient}
J.~I. Colorado-Cervantes, P.~Nardinocchi, P.~Piras, V.~Sansalone, L.~Teresi,
  C.~Torromeo, P.~E. Puddu,
  \href{https://doi.org/10.1007/s10409-021-09041-0}{Patient-specific modeling
  of left ventricle mechanics}, Acta Mechanica Sinica 38~(1) (2022) 621211.
\newblock \href {https://doi.org/10.1007/s10409-021-09041-0}
  {\path{doi:10.1007/s10409-021-09041-0}}.
\newline\urlprefix\url{https://doi.org/10.1007/s10409-021-09041-0}

\bibitem{Fortini2013Three}
S.~Fortini, G.~Querzoli, S.~Espa, A.~Cenedese,
  \href{https://doi.org/10.1007/s00348-013-1609-0}{Three-dimensional structure
  of the flow inside the left ventricle of the human heart}, Experiments in
  Fluids 54 (2013) 1--9.
\newblock \href {https://doi.org/10.1007/s00348-013-1609-0}
  {\path{doi:10.1007/s00348-013-1609-0}}.
\newline\urlprefix\url{https://doi.org/10.1007/s00348-013-1609-0}

\bibitem{DiLabbio2022Braids}
G.~Di~Labbio, J.-L. Thiffeault, L.~Kadem,
  \href{https://doi.org/10.1017/flo.2022.6}{Braids in the heart: Global
  measures of mixing for cardiovascular flows}, Flow 2 (2022) E12.
\newblock \href {https://doi.org/10.1017/flo.2022.6}
  {\path{doi:10.1017/flo.2022.6}}.
\newline\urlprefix\url{https://doi.org/10.1017/flo.2022.6}

\bibitem{Berkooz1993Proper}
G.~Berkooz, P.~Holmes, J.~L. Lumley,
  \href{https://doi.org/10.1146/annurev.fl.25.010193.002543}{The proper
  orthogonal decomposition in the analysis of turbulent flows}, Annual review
  of fluid mechanics 25~(1) (1993) 539--575.
\newline\urlprefix\url{https://doi.org/10.1146/annurev.fl.25.010193.002543}

\bibitem{Schmid2010Dynamic}
P.~J. Schmid, \href{https://doi.org/10.1017/S0022112010001217}{Dynamic mode
  decomposition of numerical and experimental data}, Journal of Fluid Mechanics
  656 (2010) 5--28.
\newblock \href {https://doi.org/10.1017/S0022112010001217}
  {\path{doi:10.1017/S0022112010001217}}.
\newline\urlprefix\url{https://doi.org/10.1017/S0022112010001217}

\bibitem{Kazemi2022Reduced}
A.~Kazemi, M.~Stoddard, A.~A. Amini,
  \href{https://doi.org/10.1117/12.2613383}{Reduced-order modeling of 4d flow
  mri and cfd in stenotic flow using proper orthogonal decomposition (pod) and
  dynamic mode decomposition (dmd)}, in: Medical Imaging 2022: Biomedical
  Applications in Molecular, Structural, and Functional Imaging, Vol. 12036,
  SPIE, 2022, pp. 509--519.
\newline\urlprefix\url{https://doi.org/10.1117/12.2613383}

\bibitem{Habibi2020Data}
M.~Habibi, S.~T. Dawson, A.~Arzani,
  \href{https://doi.org/10.3390/fluids5030111}{Data-driven pulsatile blood flow
  physics with dynamic mode decomposition}, Fluids 5~(3) (2020) 111.
\newline\urlprefix\url{https://doi.org/10.3390/fluids5030111}

\bibitem{Wu2023Flow}
X.~Wu, H.~Saaid, J.~Voorneveld, T.~Claessens, J.~J. Westenberg, N.~de~Jong,
  J.~G. Bosch, S.~Kenjere{\v{s}},
  \href{https://doi.org/10.1007/s13239-023-00684-0}{4d flow patterns and
  relative pressure distribution in a left ventricle model by shake-the-box and
  proper orthogonal decomposition analysis}, Cardiovascular Engineering and
  Technology 14~(6) (2023) 743--754.
\newline\urlprefix\url{https://doi.org/10.1007/s13239-023-00684-0}

\bibitem{Borja2024Deriving}
M.~G. Borja, P.~Martinez-Legazpi, C.~Nguyen, O.~Flores, A.~M. Kahn, J.~Bermejo,
  J.~C. Del~{\'A}lamo,
  \href{https://doi.org/10.1016/j.compbiomed.2024.108760}{Deriving
  phenotype-representative left ventricular flow patterns by reduced-order
  modeling and classification}, Computers in Biology and Medicine 179 (2024)
  108760.
\newline\urlprefix\url{https://doi.org/10.1016/j.compbiomed.2024.108760}

\bibitem{Groun2022Higher}
N.~Groun, M.~Villalba-Orero, E.~Lara-Pezzi, E.~Valero, J.~Garicano-Mena,
  S.~Le~Clainche,
  \href{https://doi.org/10.1016/j.compbiomed.2022.105384}{Higher order dynamic
  mode decomposition: From fluid dynamics to heart disease analysis}, Computers
  in Biology and Medicine 144 (2022) 105384.
\newline\urlprefix\url{https://doi.org/10.1016/j.compbiomed.2022.105384}

\bibitem{LeClainche2017Higher}
S.~Le~Clainche, J.~M. Vega, \href{https://doi.org/10.1137/15M1054924}{Higher
  order dynamic mode decomposition}, SIAM Journal on Applied Dynamical Systems
  16~(2) (2017) 882--925.
\newblock \href {https://doi.org/10.1137/15M1054924}
  {\path{doi:10.1137/15M1054924}}.
\newline\urlprefix\url{https://doi.org/10.1137/15M1054924}

\bibitem{Bell2025Automatic}
A.~Bell-Navas, N.~Groun, M.~Villalba-Orero, E.~Lara-Pezzi, J.~Garicano-Mena,
  S.~Le~Clainche, \href{https://doi.org/10.1016/j.eswa.2024.125849}{Automatic
  cardiac pathology recognition in echocardiography images using higher order
  dynamic mode decomposition and a vision transformer for small datasets},
  Expert Systems with Applications 264 (2025) 125849.
\newblock \href {https://doi.org/10.1016/j.eswa.2024.125849}
  {\path{doi:10.1016/j.eswa.2024.125849}}.
\newline\urlprefix\url{https://doi.org/10.1016/j.eswa.2024.125849}

\bibitem{Corrochano2020Flow}
A.~Corrochano, D.~Xavier, P.~Schlatter, R.~Vinuesa, S.~Le~Clainche,
  \href{https://doi.org/10.3390/fluids6010004}{Flow structures on a planar food
  and drug administration (fda) nozzle at low and intermediate reynolds
  number}, Fluids 6~(1) (2020) 4.
\newline\urlprefix\url{https://doi.org/10.3390/fluids6010004}

\bibitem{Lazpita2022Generation}
E.~Lazpita, {\'A}.~Mart{\'\i}nez-S{\'a}nchez, A.~Corrochano, S.~Hoyas,
  S.~Le~Clainche, R.~Vinuesa, \href{https://doi.org/10.1063/5.0088305}{On the
  generation and destruction mechanisms of arch vortices in urban fluid flows},
  Physics of Fluids 34~(5) (2022).
\newline\urlprefix\url{https://doi.org/10.1063/5.0088305}

\bibitem{Munoz2022Topology}
E.~Mu{\~n}oz, S.~Le~Clainche, \href{https://doi.org/10.1063/5.0080834}{On the
  topology patterns and symmetry breaking in two planar synthetic jets},
  Physics of Fluids 34~(2) (2022).
\newline\urlprefix\url{https://doi.org/10.1063/5.0080834}

\bibitem{LeClainche2018Reduced}
S.~Le~Clainche, E.~Ferrer, \href{https://doi.org/10.3390/en11030566}{A reduced
  order model to predict transient flows around straight bladed vertical axis
  wind turbines}, Energies 11~(3) (2018) 566.
\newline\urlprefix\url{https://doi.org/10.3390/en11030566}

\bibitem{Beltran2022Adaptive}
V.~Beltr{\'a}n, S.~Le~Clainche, J.~M. Vega,
  \href{https://doi.org/10.1007/s10915-022-01855-2}{An adaptive data-driven
  reduced order model based on higher order dynamic mode decomposition},
  Journal of Scientific Computing 92~(1) (2022) 12.
\newline\urlprefix\url{https://doi.org/10.1007/s10915-022-01855-2}

\bibitem{Tagliabue2017Complex}
A.~Tagliabue, L.~Ded{\`e}, A.~Quarteroni,
  \href{https://doi.org/10.1063/1.5002120}{Complex blood flow patterns in an
  idealized left ventricle: a numerical study}, Chaos: An Interdisciplinary
  Journal of Nonlinear Science 27~(9) (2017).
\newline\urlprefix\url{https://doi.org/10.1063/1.5002120}

\bibitem{Zheng2012Computational}
X.~Zheng, J.~Seo, V.~Vedula, T.~Abraham, R.~Mittal,
  \href{https://doi.org/10.1016/j.euromechflu.2012.03.002}{Computational
  modeling and analysis of intracardiac flows in simple models of the left
  ventricle}, European Journal of Mechanics - B/Fluids 35 (2012) 31--39.
\newblock \href {https://doi.org/10.1016/j.euromechflu.2012.03.002}
  {\path{doi:10.1016/j.euromechflu.2012.03.002}}.
\newline\urlprefix\url{https://doi.org/10.1016/j.euromechflu.2012.03.002}

\bibitem{Lazpita2024Modeling}
E.~Lazpita, A.~Mares, P.~Quintero, J.~Garicano-Mena, S.~Le~Clainche,
  \href{https://doi.org/10.1016/j.rineng.2024.103644}{Modeling heart flow
  dynamics using numerical simulations to identify the vortex ring: a practical
  guide}, Results in Engineering 24 (2024) 103644.
\newblock \href {https://doi.org/10.1016/j.rineng.2024.103644}
  {\path{doi:10.1016/j.rineng.2024.103644}}.
\newline\urlprefix\url{https://doi.org/10.1016/j.rineng.2024.103644}

\bibitem{Vedula2014Computational}
V.~Vedula, S.~Fortini, J.~Seo, G.~Querzoli, R.~Mittal,
  \href{https://doi.org/10.1007/s00162-014-0335-4}{Computational modeling and
  validation of intraventricular flow in a simple model of the left ventricle},
  Theoretical and Computational Fluid Dynamics 28 (2014) 589--604.
\newblock \href {https://doi.org/10.1007/s00162-014-0335-4}
  {\path{doi:10.1007/s00162-014-0335-4}}.
\newline\urlprefix\url{https://doi.org/10.1007/s00162-014-0335-4}

\bibitem{Kjeldsberg2023Verified}
H.~A. Kjeldsberg, J.~Sundnes, K.~Valen-Sendstad,
  \href{https://doi.org/10.1002/cnm.3703}{A verified and validated moving
  domain computational fluid dynamics solver with applications to
  cardiovascular flows}, International Journal for Numerical Methods in
  Biomedical Engineering 39~(6) (2023) e3703.
\newline\urlprefix\url{https://doi.org/10.1002/cnm.3703}

\bibitem{Hunt1988Turbulent}
J.~Hunt, S.~Leibovich, K.~Richards,
  \href{https://doi.org/10.1002/qj.49711448405}{Turbulent shear flows over low
  hills}, Quarterly Journal of the Royal Meteorological Society 114~(484)
  (1988) 1435--1470.
\newline\urlprefix\url{https://doi.org/10.1002/qj.49711448405}

\bibitem{He2022Numerical}
G.~He, L.~Han, J.~Zhang, A.~Shah, D.~J. Kaczorowski, B.~P. Griffith, Z.~Wu,
  \href{https://doi.org/10.1080/10255842.2021.1984433}{Numerical study of the
  effect of lvad inflow cannula positioning on thrombosis risk}, Computer
  methods in biomechanics and biomedical engineering 25~(8) (2022) 852--860.
\newline\urlprefix\url{https://doi.org/10.1080/10255842.2021.1984433}

\bibitem{Lazpita2024ECCOMAS}
E.~Lazpita, M.~Nagargoje, A.~Mares, P.~Quintero, M.~Neidlin, S.~Clainche,
  J.~Garicano-Mena,
  \href{https://www.scipedia.com/public/Lazpita_et_al_2024a}{On the numerical
  simulation of left ventricle models}, in: ECCOMAS, 2024.
\newline\urlprefix\url{https://www.scipedia.com/public/Lazpita_et_al_2024a}

\bibitem{Fluent}
Ansys, Inc., Ansys Fluent Theory Guide, Ansys Fluent Academic Research, Release
  2023 R1 (2023).

\bibitem{Starccm}
Siemens Industries Digital Software, Simcenter STAR-CCM+, version 2023.06
  (2023).

\bibitem{Sirovich1987Turbulence}
L.~Sirovich,
  \href{https://www.ams.org/journals/qam/1987-45-03/S0033-569X-1987-0910462-6/}{Turbulence
  and the dynamics of coherent structures. i. coherent structures}, Quarterly
  of Applied Mathematics 45~(3) (1987) 561--571.
\newblock \href {https://doi.org/10.1090/qam/910462}
  {\path{doi:10.1090/qam/910462}}.
\newline\urlprefix\url{https://www.ams.org/journals/qam/1987-45-03/S0033-569X-1987-0910462-6/}

\bibitem{Hetherington2024Modelflows}
A.~Hetherington, A.~Corrochano, R.~Abad{\'\i}a-Heredia, E.~Lazpita,
  E.~Mu{\~n}oz, P.~D{\'\i}az, E.~Maiora, M.~L{\'o}pez-Mart{\'\i}n,
  S.~Le~Clainche,
  \href{https://doi.org/10.1016/j.cpc.2024.109217}{Modelflows-app: data-driven
  post-processing and reduced order modelling tools}, Computer Physics
  Communications 301 (2024) 109217.
\newline\urlprefix\url{https://doi.org/10.1016/j.cpc.2024.109217}

\bibitem{Lazpita2025Efficient}
E.~Lazpita, J.~Garicano-Mena, S.~Le~Clainche, Efficient reduced order model
  based on hodmd to predict intraventricular flow dynamics, arXiv preprint
  (2025).

\bibitem{Begiashvili2023Data}
B.~Begiashvili, N.~Groun, J.~Garicano-Mena, S.~Le~Clainche, E.~Valero,
  \href{https://doi.org/10.1063/5.0142102}{Data-driven modal decomposition
  methods as feature detection techniques for flow problems: A critical
  assessment}, Physics of Fluids 35~(4) (2023).
\newline\urlprefix\url{https://doi.org/10.1063/5.0142102}

\bibitem{ModelFLOWsCardiac}
Digitheart: New tools and models for predicting heart disease progression and
  treatment response. modelflows-cardiac,
  \url{https://modelflows.github.io/modelflowsapp/cardiacpathologydetection/}.

\end{thebibliography}

\end{document}